\documentclass[letterpaper,journal,twoside]{IEEEtranTCOM}

\makeatletter
\def\endthebibliography{%
  \def\@noitemerr{\@latex@warning{Empty `thebibliography' environment}}%
  \endlist
}
\makeatother
\ifCLASSINFOpdf
\else
\fi
\usepackage{cite}
\usepackage{graphicx}
\usepackage{hyperref}
\usepackage{amsmath}
\usepackage{amssymb}
\usepackage{bm}
\usepackage{bbm}
\usepackage{algorithmicx}
\usepackage{algorithm}
\usepackage{algpseudocode}
\usepackage{array}
\usepackage{booktabs}
\usepackage{multirow}
\usepackage{makecell}
\newcommand{\C}{\bm{c}}

\newcommand{\E}{\mathbb{E}}
\newcommand{\R}{\mathbb{R}}
\newcommand{\x}{\tilde{x}}
\newcommand{\N}{\tilde{N}}

\DeclareMathOperator{\dfree}{\mathit{d}_{free}}
\DeclareMathOperator{\dcrc}{\mathit{d}_{CRC}}
\DeclareMathOperator{\Prob}{P}
\DeclareMathOperator{\Qfun}{Q}

\DeclareMathOperator{\LVA}{LVA}

\DeclareMathOperator{\CRC}{CRC}
\DeclareMathOperator{\NACK}{NACK}

\DeclareMathOperator{\UE}{UE}
\DeclareMathOperator{\var}{var}
\DeclareMathOperator{\Fail}{F}
\DeclareMathOperator{\LLB}{LLB}
\DeclareMathOperator{\NNLB}{NNLB}
\DeclareMathOperator{\NNUB}{NNUB}
\DeclareMathOperator{\RCU}{rcu}
\DeclareMathOperator{\Viterbi}{Viterbi}

\newtheorem{theorem}{Theorem}
\newtheorem{corollary}{Corollary}
\newtheorem{conjecture}{Conjecture}

\begin{document}
%
% paper title
% Titles are generally capitalized except for words such as a, an, and, as,
% at, but, by, for, in, nor, of, on, or, the, to and up, which are usually
% not capitalized unless they are the first or last word of the title.
% Linebreaks \\ can be used within to get better formatting as desired.
% Do not put math or special symbols in the title.
\title{Joint Design of Convolutional Code and CRC under Serial List Viterbi Decoding}
%
%
% author names and IEEE memberships
% note positions of commas and nonbreaking spaces ( ~ ) LaTeX will not break
% a structure at a ~ so this keeps an author's name from being broken across
% two lines.
% use \thanks{} to gain access to the first footnote area
% a separate \thanks must be used for each paragraph as LaTeX2e's \thanks
% was not built to handle multiple paragraphs
%

\author{Hengjie~Yang,~\IEEEmembership{Student Member,~IEEE,}
        Ethan~Liang,~\IEEEmembership{Student Member,~IEEE} and
        Richard D. Wesel,~\IEEEmembership{Senior Member,~IEEE}% <-this % stops a space
\thanks{This research is supported in part by National Science Foundation (NSF) grant CCF-1618272. Any opinions, findings, and conclusions or recommendations expressed in this material are those of the author(s) and do not necessarily reflect the views of the NSF. Portions of this work  will be  presented at the 2018 IEEE Global Communications Conference, Abu Dhabi, UAE \cite{Yang2018}  . }
\thanks{The authors are with Department of Electrical and Computer Engineering, University of California, Los Angeles, Los Angeles, CA 90095, USA (e-mail: \{hengjie.yang, emliang, wesel\}@ucla.edu).}% <-this % stops a space
% \thanks{J. Doe and J. Doe are with Anonymous University.}% <-this % stops a space
% \thanks{Manuscript received April 19, 2005; revised August 26, 2015.}
}

% note the % following the last \IEEEmembership and also \thanks - 
% these prevent an unwanted space from occurring between the last author name
% and the end of the author line. i.e., if you had this:
% 
% \author{....lastname \thanks{...} \thanks{...} }
%                     ^------------^------------^----Do not want these spaces!
%
% a space would be appended to the last name and could cause every name on that
% line to be shifted left slightly. This is one of those "LaTeX things". For
% instance, "\textbf{A} \textbf{B}" will typeset as "A B" not "AB". To get
% "AB" then you have to do: "\textbf{A}\textbf{B}"
% \thanks is no different in this regard, so shield the last } of each \thanks
% that ends a line with a % and do not let a space in before the next \thanks.
% Spaces after \IEEEmembership other than the last one are OK (and needed) as
% you are supposed to have spaces between the names. For what it is worth,
% this is a minor point as most people would not even notice if the said evil
% space somehow managed to creep in.

% The paper headers
\markboth{IEEE Transactions on Communications}%
{Yang \MakeLowercase{\textit{et al.}}: Joint Design of Convolutional Code and CRC under Serial List Viterbi Decoding}
% The only time the second header will appear is for the odd numbered pages
% after the title page when using the twoside option.
% 
% *** Note that you probably will NOT want to include the author's ***
% *** name in the headers of peer review papers.                   ***
% You can use \ifCLASSOPTIONpeerreview for conditional compilation here if
% you desire.

% If you want to put a publisher's ID mark on the page you can do it like
% this:
%\IEEEpubid{0000--0000/00\$00.00~\copyright~2015 IEEE}
% Remember, if you use this you must call \IEEEpubidadjcol in the second
% column for its text to clear the IEEEpubid mark.

% use for special paper notices
%\IEEEspecialpapernotice{(Invited Paper)}

% make the title area
\maketitle

% As a general rule, do not put math, special symbols or citations
% in the abstract or keywords.
\begin{abstract}
This paper studies the joint design of optimal convolutional codes (CCs) and CRC codes when serial list Viterbi algorithm (S-LVA) is employed in order to achieve the target frame error rate (FER). We first analyze the S-LVA performance with respect to SNR and list size, repsectively, and prove the convergence of the expected number of decoding attempts when SNR goes to the extreme. We then propose the coded channel capacity as the criterion to jointly design optimal CC-CRC pair and optimal list size and show that the optimal list size of S-LVA is always the cardinality of all possible CCs. With the maximum list size, we choose the design metric of optimal CC-CRC pair as the SNR gap to random coding union (RCU) bound and the optimal CC-CRC pair is the one that achieves a target SNR gap with the least complexity. Finally, we show that a weaker CC with a strong optimal CRC code could be as powerful as a strong CC with no CRC code.
\end{abstract}

% Note that keywords are not normally used for peerreview papers.
\begin{IEEEkeywords}
Convolutional code, cyclic redundancy check (CRC) code, serial list Viterbi algorithm (S-LVA), coded channel capacity, random coding union (RCU) bound
\end{IEEEkeywords}

% For peer review papers, you can put extra information on the cover
% page as needed:
% \ifCLASSOPTIONpeerreview
% \begin{center} \bfseries EDICS Category: 3-BBND \end{center}
% \fi
%
% For peerreview papers, this IEEEtran command inserts a page break and
% creates the second title. It will be ignored for other modes.
\IEEEpeerreviewmaketitle

\section{Introduction}
% The very first letter is a 2 line initial drop letter followed
% by the rest of the first word in caps.
% 
% form to use if the first word consists of a single letter:
% \IEEEPARstart{A}{demo} file is ....
% 
% form to use if you need the single drop letter followed by
% normal text (unknown if ever used by the IEEE):
% \IEEEPARstart{A}{}demo file is ....
% 
% Some journals put the first two words in caps:
% \IEEEPARstart{T}{his demo} file is ....
% 
% Here we have the typical use of a "T" for an initial drop letter
% and "HIS" in caps to complete the first word.
\IEEEPARstart{C}{yclic} redundancy check (CRC) codes \cite{Blahut2003} are commonly used as the outer error-detection code for an inner error-correction code. An undetected error (UE) occurs when the erroneously decoded sequence passes the CRC check.

In a convolutionally encoded system, the list Viterbi decoding algorithm (LVA) produces an ordered list of decoded sequences in order to decode beyong the free distance of the convolutional code. For serial LVA (S-LVA), the algorithm terminates when a decoded sequence passes the CRC check or the list size has been exhausted. 

With a target frame error rate (FER), this paper aims at designing the optimal convolutional code and the optimal CRC code, i.e., the optimal CC-CRC pair, to achieve the target FER with the least possible decoding complexity of S-LVA.

\subsection{Previous Work}
In \cite{PK2004}, Koopman and Chakravarty list the commonly used CRC codes up to degree 16.  The designs in  \cite{PK2004}  as with most CRC designs, assume that the CRC decoder operates on a binary symmetric channel (BSC), whereas in reality the CRC decoder sees message sequences whose likelihoods depend on the codeword structure of the inner code. 

For an inner convolutional code (CC), Lou \emph{et al.} \cite{CY2015}, for the first time, studied the design of a CRC code specifically for the inner CC. The authors presented two methods to obtain an upper bound on the UE probability of any CC-CRC pair. These methods were called the exclusion method and the construction method. A greedy CRC code search algorithm was proposed by using the fact that when FER is low, UEs with the smallest Hamming distance dominate performance. Using this search algorithm, the authors in \cite{CY2015} obtained the ``distance-spectrum-optimal'' CRC codes that minimize the UE probability, $\Prob_{\UE}$. Here, a distance-spectrum-optimal CRC code refers to a CRC code that maximizes the distance between arbitrarily two different CCs. As an example, for a commonly used 64-state CC with 1024 information bits, the distance-spectrum-optimal CRC code typically requires 2 fewer bits to achieve a target $\Prob_{\UE}$ or to reduce the $\Prob_{\UE}$ by orders of magnitude (at high SNR) over the performance of standard CRC codes with the same degree.

The list Viterbi algorithm (LVA) \cite{Johannesson1999} produces an ordered list of the $L$ most likely transmitted codewords. Parallel LVA produces these $L$ codewords all at once. Serial LVA (S-LVA) produces codewords one at a time until the CRC check passes; see Seshadri and Sundberg \cite{SS1994}. Several implementations of fast LVAs have appeared in literature \cite{Soong1991,SS1994,Nill1995,Roder2006}. Soong and Huang \cite{Soong1991} proposed an efficient tree-trellis algorithm (TTA), which is a serial LVA, initially used for speech recognition. Roder and Hamzaoui \cite{Roder2006} then improved the TTA by using several unsorted lists to eventually provide the list of $L$ best sequences, allowing the TTA to achieve linear time complexity with respect to the list size. Wang \emph{et al.} \cite{Wang2008} proposed using the parity-check matrix of the CRC generator polynomial to assist decoding in a convolutionally coded system. If the soft Viterbi decoding fails, the CRC-CC pair is jointly decoded iteratively until a codeword passes the CRC check. As for complexity, Sybis \emph{et al.} \cite{Sybis2016} presented a table which quantifies the complexity cost for basic operations, such as addition, multiplication, division, comparision and table look-up operations and provided detailed complexity calculation for various codes in moderate blocklength.

Despite the different implementations of LVA, several literatures \cite{Chen2001,Bai2004,Lijofi2004} also study different variations of LVA. Chen and Sundberg \cite{Chen2001} studied the LVA for continuous transmission using tail-biting CC and proved that as $L$ increases, the LVA asymptotically approaches the pure maximum likely (ML) error correction decoder, which is referred to as asymptotic optimality. Bai \emph{et al.} \cite{Bai2004} analyzed the performance and arithmetic complexity of parallel concatenated convolutional codes. For S-LVA, Lijofi \emph{et al.} \cite{Lijofi2004} proposed a list single-wrong turn (SWT) convolutional decoding algorithm that is computationally less complex than S-LVA. Instead of choosing the $L$ most likely paths, the list-SWT Viterbi algorithm determines $L$ paths that are direct descendents of the best path. Despite the suboptimality of list SWT Viterbi algorithm, it achieves nearly the same BER and FER performance of S-LVA under Gaussian channel and Rayleigh channel. 

In the finite blocklength regime, Polyanskiy \emph{et al.} \cite{Polyanskiy2010} studied the fundamental channel coding rate, in which the average probability of error $\epsilon$ for the best $(n,M,\epsilon)$ code is upper bounded by the random coding union (RCU) bound $\RCU(n,M)$. This bound is seen as a benchmark for a practical code used in finite blocklength. However, the computation of RCU bounds involves integrating $n$-dimensional vectors, which is computationally prohibitive even for moderate values of $n$. Font-Segura \emph{et al.} \cite{Segura2018} proposed a saddlepoint method to simplify the computation of RCU bound.

\subsection{Main Contributions}\label{sec: main contributions}
In this paper, we consider the design problem of finding the optimal CC-CRC pair when S-LVA decoder is employed to achieve the target FER with the least possible decoding complexity. The candidate CC-CRC pairs considered in this paper are the ones of a most popular CC in \cite{ErrorControlCoding} used with a distance-spectrum-optimal CRC code designed using Lou \emph{et al.}'s method \cite{CY2015}. First, we model the system as a \emph{coded channel} that consists of the CRC encoder, the convolutional encoder, the AWGN channel, the S-LVA decoder and the CRC decoder, which, as a whole, can be seen as an error and erasure channel. In parallel with the classical definition of the channel capacity, the \emph{coded channel capacity} is the maximum bits per codeword transmission. With the target FER, the optimal CC-CRC pair with the optimal list size of S-LVA should maximize the coded channel capacity. Since the design of list size $L$ is independent of the design of CC-CRC pair, we show that $L=|\mathcal{C}|$ is always the optimal list size for any candidate CC-CRC pair.  With $L=|\mathcal{C}|$ fixed, since all CC-CRC pairs that could achieve the target FER have roughly the same coded channel capacity, we choose the design metric as the SNR gap to RCU bound and the optimal CC-CRC pair is the one that has the target SNR gap with the least decoding complexity.

In the coded channel model, the S-LVA combined with the optimal CRC code designed using \cite{CY2015} specifically for a given CC is of significant interest as well. We will first study the decoding performance of S-LVA in order to provide the reader with a better understand of properties of the probability of error and probability of erasure. 

In summary, the main contributions of this paper are as follows.
\begin{enumerate}
\item Since the list size $L$ determines the maximum number of codewords the S-LVA will check and $L$ ranges from $1$ to $|\mathcal{C}|$, where $\mathcal{C}$ is the set of all possible convolutional codes, this paper uses bounds, approximations, and simulations to characterize the trade-off between two probabilities: the erasure probability $\Prob^{L}_{\NACK}$, when no codeword passes the CRC check producing a negative acknowledgement (NACK) and the UE probability $\Prob^{L}_{\UE}$ when an incorrect codeword passes the CRC.

\item The complexity of S-LVA is captured by the expected number of decoding attempts. For S-LVA with a degree-$m$ CRC code and the maximum possible list size $L=|\mathcal{C}|$, we first prove that the expected number of decoding attempts converges to $2^m(1-\epsilon)$, for a small $\epsilon>0$, as SNR decreases and to $1$ as SNR increases. We also propose the time ratio of traceback or insertion to a standard Viterbi operation as the complexity metric and give the analytical expression to evaluate the empirical time complexity.

\item We first propose the coded channel capacity as a useful criterion to select the optimal CC-CRC pair and list size $L$. We show that the best performance for any CC-CRC pair is always attained when $L=|\mathcal{C}|$, regardless of SNR. With $L=|\mathcal{C}|$ fixed, we choose the SNR gap to RCU bound as the design metric of finding the optimal CC-CRC pair. We also provide sufficient evidences to show that a weaker CC used with a stronger CRC code can achieve nearly the same performance as a single strong CC with no CRC code.
\end{enumerate}

\subsection{Organization}
This paper is organized as follows. Section \ref{sec:system model} introduces the system model. Section \ref{sec:S-LVA} analyzes the decoding performance and complexity and proves the convergence of the expected number of decoding attempts. Section \ref{sec:coded channel} describes the coded channel model and several simplified models. Section \ref{sec:optimal CC-CRC} presents the design methodology and design examples of the optimal CC-CRC pair to achieve the target FER among all candidate CC-CRC pairs. Section \ref{sec:conclusion} concludes the paper. 

\begin{figure}[t]
\centering
\includegraphics[scale=0.53]{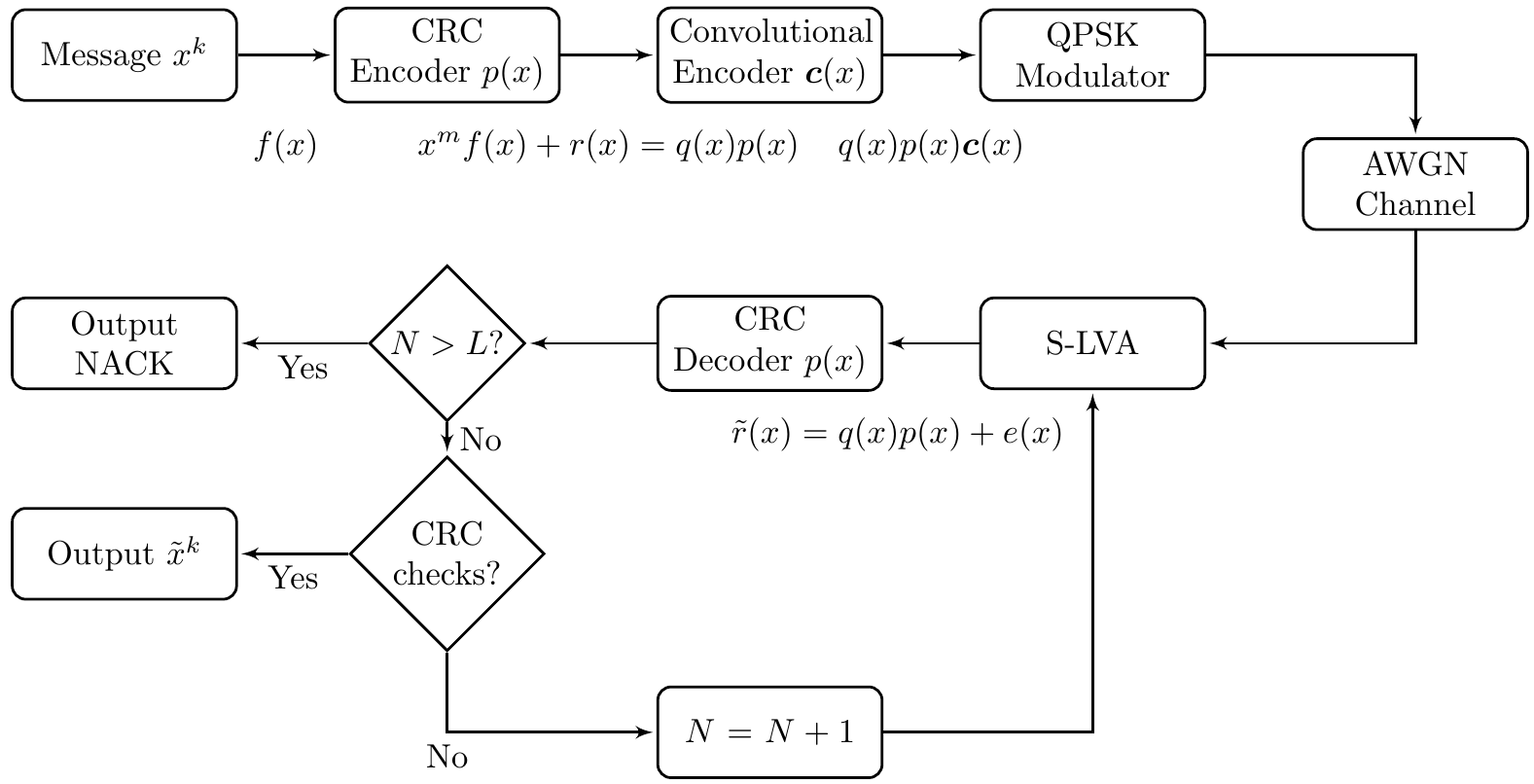}
\caption{Block diagram of a system employing convolutional codes, CRC codes and S-LVA decoder}
\label{fig:system model}
\end{figure}

\section{System Model}\label{sec:system model}

The system model we study in this paper is shown in Fig. \ref{fig:system model}. A transmitter uses a CC and a CRC code to transmit an information sequence as follows: Let $f(x)$ denote a $k$-bit binary information sequence and $p(x)$ denote a degree-$m$ CRC generator polynomial. Let $r(x)$ denote the remainder when $x^mf(x)$ is divided by $p(x)$. First, the CRC polynomial is used to obtain the $n=k+m$-bit sequence $x^m f(x) + r(x) = q(x)p(x)$. The transmitter then uses a feedforward, rate-$\frac1{N}$ CC with $v$ memory elements and a generator polynomial $\C(x)$ to encode the $n$-bit sequence. The output $q(x)p(x)\C(x)$ of the convolutional encoder is transmitted over an additive white Gaussian noise (AWGN) channel using quadrature phase-shift keying (QPSK) modulation.

The receiver feeds the noisy received sequence into a S-LVA decoder with list size $L$ that identifies $L$ most likely $n$-bit input sequences sequentially. That is, S-LVA begins by finding the closest codeword $c_1$ to the received sequence and passing it to the CRC code for verification. If the CRC check fails, S-LVA outputs the next closest codeword $c_2$ and repeats the above procedure until the CRC check is successful or the best $L$ codewords $c_1, \ldots c_{L}$ all fail the CRC check, in which case the decoder declares erasure and a NACK is generated.

In this paper, unless otherwise stated, the CRC code in the system model is the one designed using the CRC code search algorithm in \cite{CY2015} for the given convolutional code, in which the authors also provide the analytical upper bound on the undetected error probability with two different methods, the exclusion method and the construction method. We refer interested readers to \cite{CY2015} for more details.

\section{S-LVA Performance Analysis}\label{sec:S-LVA}
From Sec. \ref{sec:system model}, it can be seen that the failure rate of S-LVA can be expressed as
\begin{align}
\Prob^{L}_{\Fail}=\Prob^{L}_{\UE}+\Prob^{L}_{\NACK},
\end{align}
where $\Prob^{L}_{\UE}$ and $\Prob^{L}_{\NACK}$ are both a function of SNR and list size $L$. The performance metrics of S-LVA include  $\Prob^{L}_{\Fail}$, $\Prob^{L}_{\UE}$, $\Prob^{L}_{\NACK}$, and $\E[N_{\LVA}]$. In fact, $\Prob^{L}_{\UE}$ and $\Prob^{L}_{\NACK}$ reflect the overall characteristics of the coded channel model introduced in Sec. \ref{sec: main contributions} as the coded channel requires the complete knowledge of transition probabilities from the transmitted codeword to the decoded codeword or NACK. Therefore it is important to understand how the SNR and list size $L$ affect $\Prob^{L}_{\UE}$ and $\Prob^{L}_{\NACK}$, respectively.

\subsection{S-LVA Performance vs. SNR}
This section examines S-LVA performance as a function of SNR ($E_s/N_0$). The extreme cases of SNR (very low and very high) and list size ($L=1$ and $L= {| {\cal C} |}$) are given particular attention as they frame the overall performance landscape.

In the discussion below, certain sets of codewords are important to consider. First, ${\cal C}$ is the set of all convolutional codewords. Since we consider a finite blocklength system where there are $n$ message bits and $v$ termination bits (completely determined by the $n=k+m$ message bits) fed into the convolutional encoder, the size of $\mathcal{C}$ is
\begin{align} 
| {\cal C} |=2^{n} =2^{k+m}.
\end{align}
Let $c^*$ denote the transmitted codeword. A superscript of $-$ indicates a set that excludes $c^*$. For example ${\cal C}^-$ is the set of all convolutional codewords except the transmitted codeword $c^*$.
The set ${\cal C}_{\CRC}$ is the set of all convolutional codewords whose corresponding input sequences pass the CRC check. The size of ${\cal C}_{\CRC}$ is
\begin{align}
| {\cal C}_{\CRC} |=2^{n-m} =2^{k}.
\end{align}
The set ${\cal C}_{\overline{{\CRC}}}$ is the set of all convolutional codewords whose corresponding input sequences {\em do not} pass the CRC check. The size of this set is
\begin{align}
| {\cal C}_{\overline{{\CRC}}} |=2^{n} - 2^{k}.
\end{align}

\subsubsection{The Case of $L= {| {\cal C} |}$} \label{sec:L=C}
Consider S-LVA with the largest possible list size $L= {| {\cal C} |}$.  Regardless of SNR, $\Prob^{| {\cal C} |}_{\NACK}=0$ always holds because S-LVA with $L= {| {\cal C} |}$ will always find a codeword that passes the CRC check. Let $A_{d}$ be the number of distinct UEs of distance $d$ with positions taken into account. The UE probability $\Prob^{| {\cal C} |}_{\UE}$ is upper bounded by the union bound that some codeword in ${\cal C}_{\CRC}^-$ is pairwise more likely than $c^*$:
\begin{equation} \label{eq:UEUB} 
\Prob^{| {\cal C} |}_{\UE} \le \sum_{c \in {\cal C}_{\CRC}^-} \Prob (d(c,c^*)), 
\end{equation}
where $d(c,c^*)$ is the distance between $c$ and $c^*$, and $ \Prob (d(c,c^*))$ is the pairwise error probability of an error event with distance $d$. For QPSK modulation over the AWGN channel, $\Prob(d)$ can be computed using the Gaussian Q-function:
\begin{align}
\Prob(d)=\Qfun(\sqrt{d\gamma_s})\le \Qfun(\sqrt{\dfree \gamma_s})e^{-(d-\dfree)\gamma_s/2},
\end{align}
where $\gamma_s=E_s/N_0$ is the signal-to-noise ratio (SNR) of a QPSK symbol, and $E_s$ and $N_0/2$ denote the energy per transmitted QPSK symbol and one-dimensional noise variance, respectively. \footnote{In \cite{CY2015}, there is a typo in the expression for equation (2) that includes erroneously a factor of two in the square root.}. 

Here, we point out that \eqref{eq:UEUB} is precisely the union bound of \cite{CY2015} given as an upper bound on $\Prob^{1}_{\UE}$. That it is also a valid upper bound for $\Prob^{| {\cal C} |}_{\UE}$ indicates that, at least at low SNR, this bound will be loose for $L=1$.  At very low SNR, $ \Prob^{| {\cal C} |}_{\UE}$ converges to $\frac{|{\cal C}_{\CRC}^- |}{|{\cal C}_{\CRC}|} \approx 1$. We refer the reader to \cite{CY2015} for the exact expression of the union bound.

\begin{figure}[t]
\centering
\includegraphics[scale=0.4]{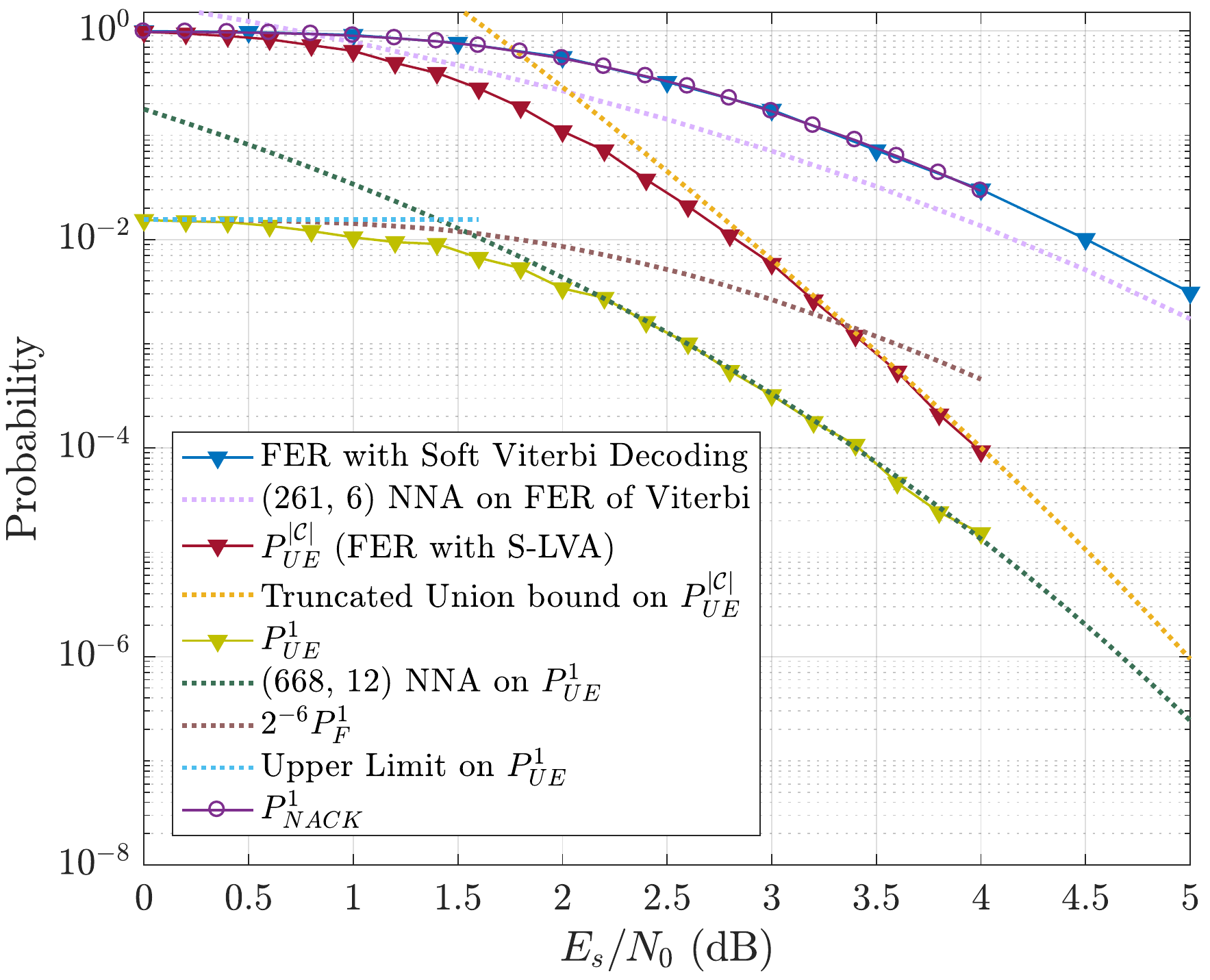}
\caption{Comparison of FER between S-LVA combined with the optimal degree-$6$ CRC code 0x43 and soft Viterbi decoding (without a CRC code) for $(13, 17)$ CC when $n=256+6$ bits. $(261,\ 6)$ NNA on soft Viterbi decoding, truncated union bound at $\tilde{d}=24$ on $\Prob_{\UE}^{|\mathcal{C}|}$, conjecture of $2^{-6}\Prob_{\Fail}^1$, upper limit of $2^{-6}$, and $(668,\ 12)$ NNA on $\Prob_{\UE}^{1}$ are also provided as a reference.}
\label{fig3}
\end{figure}

For $k=256$ bits,  Fig.~\ref{fig3} shows $\Prob^{| {\cal C} |}_{\UE}$ as a function of $E_s/N_0$ for the $(13,17)$ CC using soft Viterbi decoding without a CRC code and S-LVA with $L=| {\cal C} |$ combined with the optimal degree-$6$ CRC code 0x43. The truncated union bound at $\tilde{d}=24$ on $\Prob^{| {\cal C} |}_{\UE}$ of \eqref{eq:UEUB} derived via exclusion method in \cite{CY2015} is also shown. It can be seen that the union bound on $\Prob^{| {\cal C} |}_{\UE}$ becomes tight as SNR increases.

\subsubsection{The Case of $L= 1$}  \label{sec:L=1}

For $L=1$, with the same blocklength $n$, $\Prob^{1}_{\Fail}$ is exactly the FER of the CC under soft Viterbi decoding with no CRC code. The addition of the CRC code separates the failures into erasures and UEs, with probabilities $\Prob^{1}_{\NACK}$ and $\Prob^{1}_{\UE}$, respectively. Thus we have union bounds, nearest neighbor approximation (NNA), and a low-SNR upper limit as follows:
\begin{align} \label{eq:NACK_UB}
 \Prob^{1}_{\NACK} &\le \sum_{c \in {\cal C}_{\overline{{\CRC}}}} \Prob (d(c,c^*))\\
\label{eq:NACK_NNA} & \approx A_{\dfree} \Prob (\dfree), 
\end{align}
\begin{align}
 \label{eq:UE1_UB}  \Prob^{1}_{\UE} &\le \sum_{c \in {\cal C}_{\CRC}^-} \Prob (d(c,c^*))\\
 \label{eq:UE1_NNA}  &\approx A_{\dcrc} \Prob (\dcrc), 
\end{align} 
\begin{equation}\label{eq:upper_limit}
\lim_{\gamma_s\to-\infty}\Prob^{1}_{\UE}=2^{-m},
\end{equation}
where $A_d$ denotes the number of distinct UEs at distance $d$ with positions taken into account.

 Note that \eqref{eq:UE1_UB}  is identical to \eqref{eq:UEUB}, but $\Prob^{1}_{\UE}$ should be significantly smaller than $\Prob^{| {\cal C} |}_{\UE}$. Thus we propose an improved bound on $\Prob^{1}_{\UE}$ as follows:  for a randomly chosen degree-$m$ CRC code and $L=1$ we expect an incorrectly chosen convolutional codeword to pass the CRC check with probability $2^{-m}$. This should be an upper bound on the performance of CRCs optimized according to \cite{CY2015}.  Thus we conjecture that
 \begin{align}\label{eq:UE1conjecture} 
 \Prob^{1}_{\UE} &\le  2^{-m}  \Prob^{1}_{\Fail} \, .
\end{align}
This upper bound should be loose for well-designed CRCs at high SNR.  However, at very low SNR we expect this bound to be tight based on the fact that the upper limit of $\Prob_{\UE}^{1}$ satisfies \eqref{eq:upper_limit}.  Fig. \ref{fig3} shows that  \eqref{eq:UE1conjecture}  is accurate at very low SNR and the NNA of $\Prob_{\UE}^{1}$ in \eqref{eq:UE1_NNA} is quite accurate at high SNR.  The parameters of the NNA are $A_{\dcrc}=668$ and $\dcrc=12$.

\subsection{Complexity Analysis of S-LVA}
In \cite{Roder2006}, the authors present tables that compare the time and space complexity for different implementations of the LVA. Although the multiple-list tree-trellis algorithm (ml-TTA) achieves linear time complexity for the backward passes of the S-LVA, the implementation does not support floating point precision without the use of quanitization. The T-TTA is another implementation of the S-LVA that uses a red-black tree to store the cumulative metric differences during a traceback operation. Their time complexity results indicate that the T-TTA achieve the best performance for algorithms that support floating point precision. The analysis of the S-LVA in this assumes the use of the T-TTA.

For a fixed blocklength and a specified CC-CRC pair, the decoding complexity of S-LVA depends mainly on the number of decoding trials performed. Denote by $N_{\LVA}$ the random variable indicating the number of decoding trials of S-LVA for a received codeword randomly drawn according to the noise distribution. First, we show that with list size $|\mathcal{C}|$, the expected value of $N_{\LVA}$, $\E[N_{\LVA}]$, converges to $1$ as SNR increases and converges to $2^m(1-\epsilon)$, for a small $\epsilon > 0$ as SNR decreases. Next, we prove that $N_{\LVA}$ is a bounded random variable where the upper bound is approximately the number of all possible convolutional codes within $d_{\CRC}$. Finally, we measure the complexity of S-LVA by the time ratio, which is the ratio of the actual time an insertion or traceback operation consumes to the actual time a standard Viterbi algorithm consumes, which is the complexity of add-compare-select (ACS) operations in trellis building plus one traceback operation.

\begin{theorem}\label{thm01}
The expected number of decoding trials  $\E[N_{\LVA}]$ for S-LVA with list size $|\mathcal{C}|$, used with a degree-$m$ CRC code, satisfies (i) $\lim_{\gamma_s\to\infty}\E[N_{\LVA}]=1$; (ii) $\lim_{\gamma_s\to-\infty}\E[N_{\LVA}]=2^m(1-\epsilon)$, where $\epsilon \to 0$ as $n \to \infty$.
\end{theorem}

\begin{IEEEproof}
Let $\x_i^n$ denote the $i^{\text{th}}$ output of the S-LVA, which is the codeword at position $i$ in the list of all possible codewords sorted according to increasing soft Viterbi metric (typically Hamming or Euclidean distance) with respect to the received noisy codeword.

(i) Consider the event $A_i\triangleq\cap_{j=1}^{i-1}\{p(x)\nmid\x^n_j\}\cap\{p(x)\mid \x^n_i\},$ where $p(x)$ is the CRC polynomial.  Because of the existence of codewords that have $p(x)$ as a factor (i.e. that pass the CRC check), there exists a maximum decoding depth $\N<\infty$ such that $\Pr\{A_j\}=0, \forall j>\N$. 

Note that when $\gamma_s\to\infty$, $\Pr\{A_1\}\to1$ and $\sum_{i=2}^{\N}\Pr\{A_i\}\to0$. Thus,
\begin{align}
\lim_{\gamma_s\to\infty}\E[N_{\LVA}]&=\lim_{\gamma_s\to\infty}\left[1\cdot\Pr\{A_1\}+\sum_{i=2}^\infty i\Pr\{A_i\}\right]\notag\\
&=\lim_{\gamma_s\to\infty}\left[1\cdot\Pr\{A_1\}+\sum_{i=2}^{\N} i\Pr\{A_i\}\right]\notag\\
&\le\lim_{\gamma_s\to\infty}\left[1\cdot\Pr\{A_1\}+\N\sum_{i=2}^{\N}\Pr\{A_i\}\right]\notag\\
&=1.
\end{align}

Since $N_{\LVA} \ge 1$, $\E[N_{\LVA}] \ge 1$. It follows that $\lim_{\gamma_s\to\infty}\E[N_{\LVA}]=1$.

(ii) When $\gamma_s\to-\infty$, the SNR is low enough such that with high probability the received sequence $\bm{y}$ is far away from the entire constellation of all possible sequences that can be transmitted in $\R^n$. This implies that with very high probability $\bm{y}$ is almost equidistant from all possible convolutional codewords that can be transmitted.   For those received sequences almost equidistant from all convolutional codewords, the S-LVA decoding process can be modeled as follows: In a basket of "blue" balls (codewords that pass the CRC check) and "red" balls (codewords that do not pass the CRC check), the S-LVA chooses balls at random without replacement with the objective of stopping when it successfully picks a blue ball. 
Thus, $\E[N_{\LVA}]$ can be computed using a standard result in combinatorics as follows. For a decoded sequence with $n$ message and parity-check bits and $v$ trailing zero bits, the total number of balls in the basket is $N = 2^{n}$ and the number of blue balls in the basket is $M = 2^{n-m}$:
\begin{align}
\lim_{\gamma_s\to-\infty} \E[N_{\LVA}] &= 1 + \frac{N-M}{M + 1} \notag \\
             &= \frac{N+1}{M+1}\notag\\
             &= 2^m \left[1 - \frac{2^{m}-1}{2^{m}+2^{n}}\right]\notag\\
             &=2^m(1-\epsilon),
\end{align}
where $\epsilon=\frac{2^{m}-1}{2^{m}+2^{n}}>0$. When $m$ is fixed, $\lim_{n\to\infty}\E[N_{\LVA}]=2^m$.
\end{IEEEproof}

\begin{figure}[t]
\centering
\includegraphics[scale=0.4]{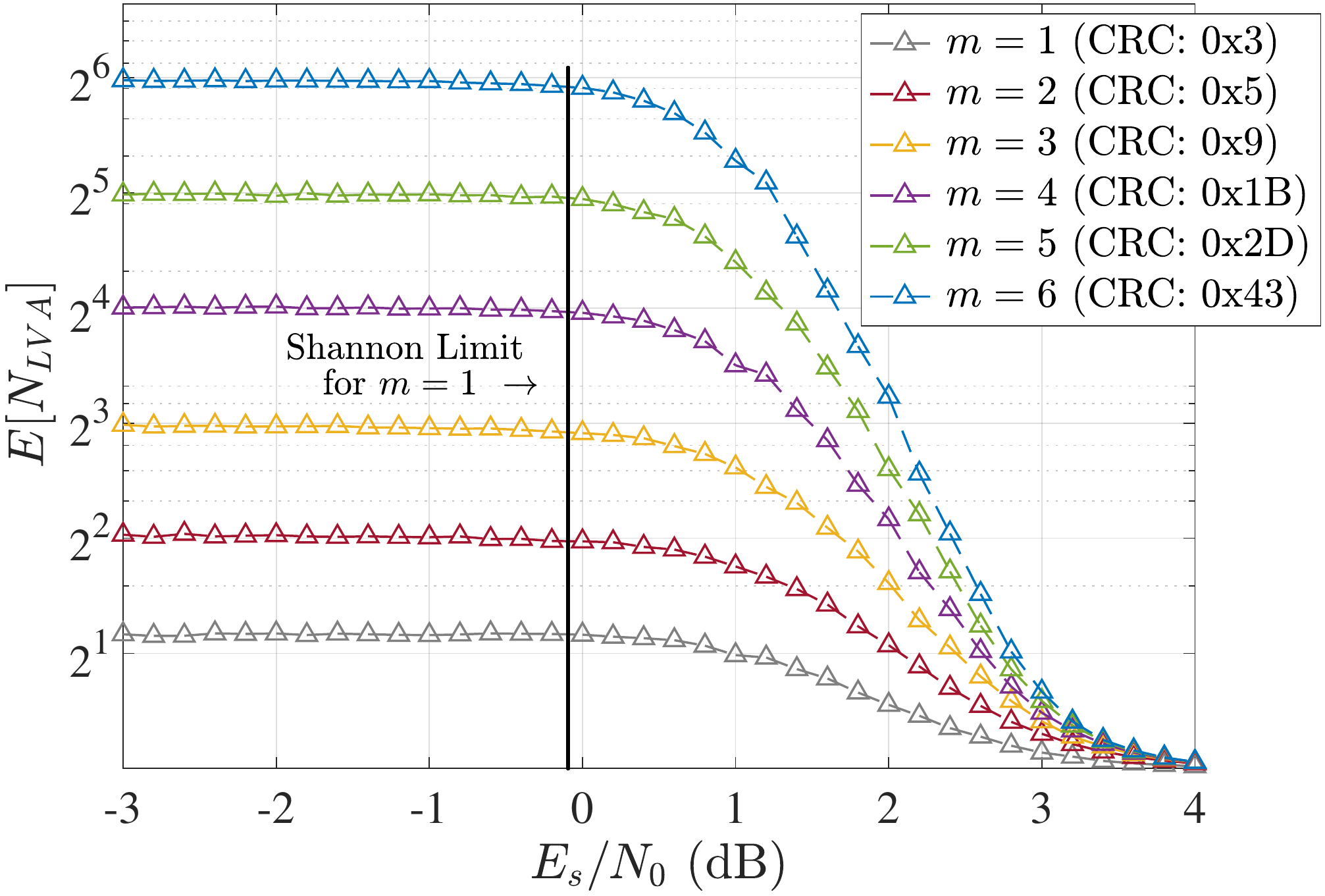}
\caption{$\E[N_{\LVA}]$ vs. $E_s/N_0$ of degree $1-6$ optimal CRC codes for $(13, 17)$ CC, with $k=256$.}
\label{fig4}
\end{figure}

Fig.~\ref{fig4} shows empirical $\E[N_{\LVA}]$ for the $(13, 17)$ CC with the optimal CRC codes with degrees ranging from $1$ to $6$ when $k=256$ bits.  The curves verify Theorem~\ref{thm01}; $\E[N_{\LVA}]\to1$ as the SNR increases and $\E[N_{\LVA}]\approx2^m$ as the SNR decreases to very low values. While the result we have obtained in Theorem \ref{thm01} for the case of $\gamma_s \to -\infty$ requires very low SNR values for the arguments made to hold, it is interesting to see from the figure that S-LVA behaves similar to random guessing as soon as the SNR value is below the Shannon limit, shown as a vertical line for $m=1$.  (The limits for the other values of $m$ are very close to the limit for $m=1$). 

Theorem \ref{thm01} studies the limit of $\E[N_{\LVA}]$ in the limit of extremely high and low SNR regimes. In practice, SNRs ranging between 0.5 dB and 4 dB above the Shannon limit are of particular interest. As shown in Fig. \ref{fig4}, $\E[N_{\LVA}]$ traverses its full range from $\approx 2^m$ to $1$ in this range of practical interest.

\begin{theorem}\label{theorem: N_LVA is bounded}
The number of decoding attempts of S-LVA with list size $L=|\mathcal{C}|$, $N_{\LVA}$, is upper bounded by
\begin{align}
N_{\LVA}\le \sum_{d=\dfree}^{d_{\CRC}}B_d-A_{d_{\CRC}}+1,
\end{align}
where $B_d$ denotes the number of all possible CCs with distance $d$, and $A_{d_{\CRC}}$ denotes the number of UEs with distance $d_{\CRC}$, both with positions taken into account.
\end{theorem}
\begin{IEEEproof}
Since the Gaussian noise is independent of the transmitted codeword, the all-zero CC can always be thought of as the transmitted CC and the surrounding CCs are the error events. Since all-zero message sequence can already pass the CRC check. The upper bound can be obtained by finding the maximum number of codewords until S-LVA finds the second CC whose input sequence can pass the CRC check.

Now consider the following extreme case: First, if S-LVA decode $S$ times, where $S=\sum_{d=\dfree}^{d_{\CRC}}B_d$, it certainly can hit a CC whose input codeword checks the CRC, since $S$ trials will include the undetectable nearest neighbors of all-zero CC. Note that here, the undetectable nearest neighbors are the relative constellation points of the true nearest neighbors of the transmitted CC. Thus by subtracting the number of undetectable nearest neighbors and then adding back one undetectable nearest neighbor, we know that the S-LVA will terminate as well by decoding at most $S-N+1$ times, which concludes that $S-N+1$ is a valid upper bound.
\end{IEEEproof}

\begin{figure}[t]
\centering
\includegraphics[scale=0.47]{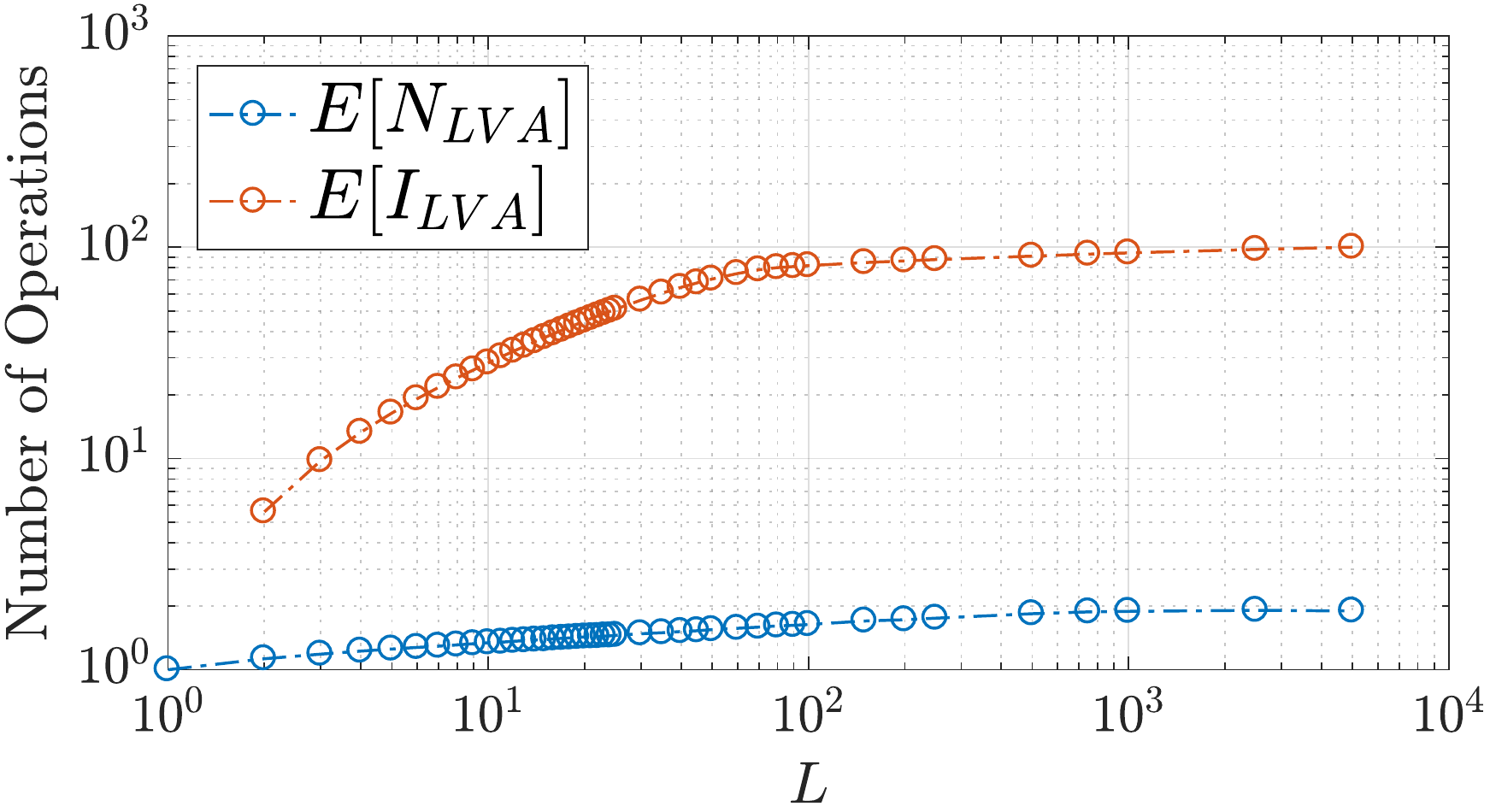}
\caption{The expected number $\E[N_{\LVA}]$ of decoding attempts and expected number $\E[I_{\LVA}]$ of insertions performed with different list sizes for $(27,31)$ CC, and 0x709 CRC code, with $k=64$ at $2$ dB. In the simulaiton setting, $C_1=1.5$ and $C_2=2.2$.}
\label{fig: expected decoding and insertion}
\end{figure}

\begin{figure}[t]
\centering
\includegraphics[scale=0.35]{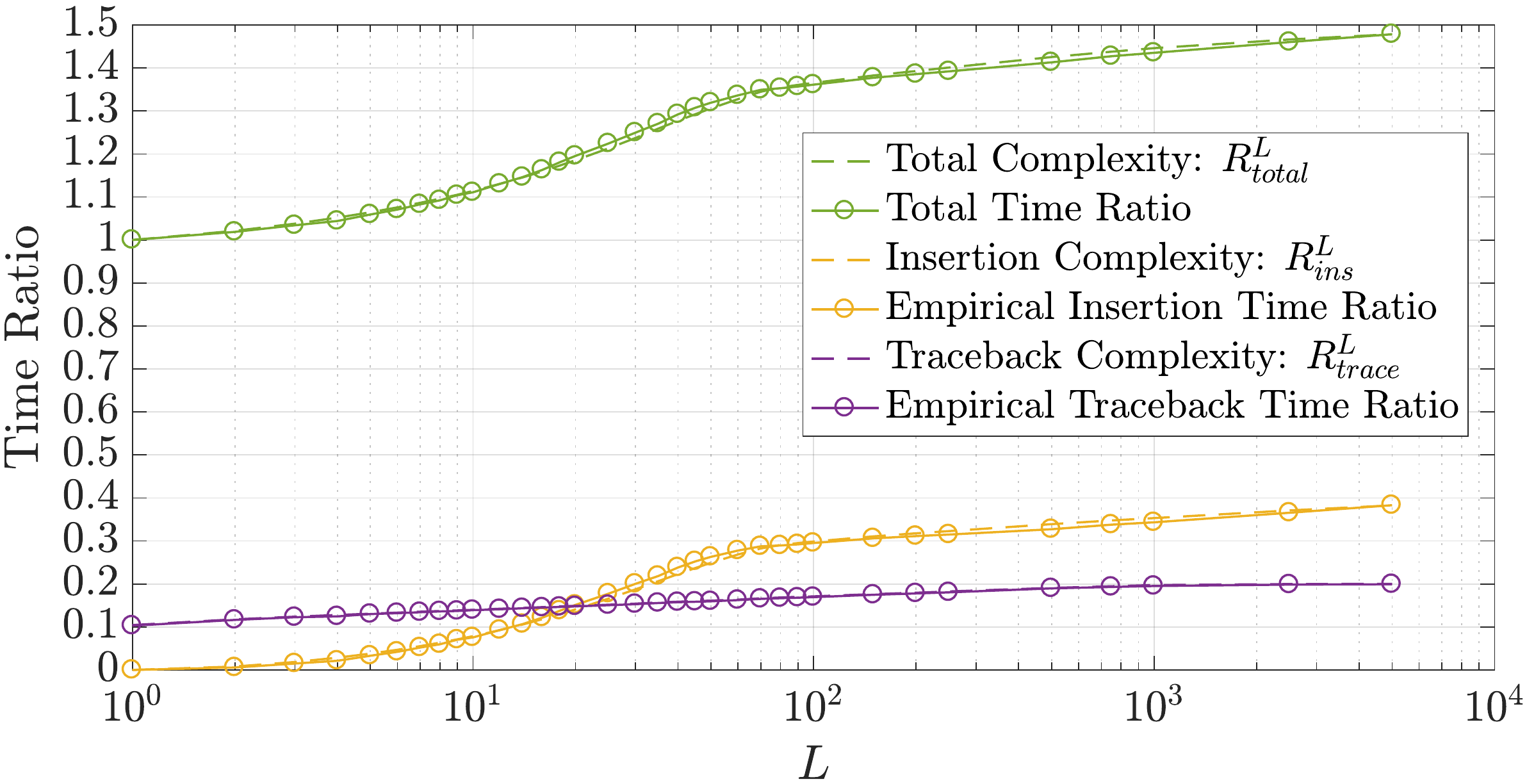}
\caption{The complexity of S-LVA with different list sizes for $(27,31)$ CC, and 0x709 CRC code, with $k=64$ at $2$ dB. In the simulaiton setting, $C_1=1.5$ and $C_2=2.2$.}
\label{fig: complexity measurement}
\end{figure}

Theorem \ref{theorem: N_LVA is bounded} shows that the number of decoding attempts of S-LVA is a bounded random variable, which means that it is enough to set list size $L=\sum_{d=\dfree}^{d_{\CRC}}B_d-A_{d_{\CRC}}+1$ which is far less than $|\mathcal{C}|$. 

Although the complexity of S-LVA is determined by $\E[N_{\LVA}]$, still, it would be interesting to investigate how time complexity changes as list size $L$ varies. Here, we define the complexity metric of S-LVA as the \emph{time ratio} $R^L_{total}$, which is the ratio of the actual time an insertion or traceback operation consumes to the actual time a standard Viterbi algorithm consumes. This metric provides a quantititive measure on the time consumption any other steps in the algorithm would cost compared to that of a standard Viterbi algorithm. 

Note that S-LVA mainly comprises two steps: an ACS operation and multiple tracebacks where the multiple tracebacks require a dynamic sorted list to obtain the next position of detour state on trellis. Thus, the time complexity of multiple tracebacks can be further split into the complexity of obtaining one trellis path and the complexity of insertions required to maintain the sorted list. When list size is large, both complexities can be seen as independent. 

Let $R_{trace}^L$ denote the time ratio of retrieving a single trellis path and $R_{ins}^L$ denote the time ratio of insertions, we have
\begin{align}
R_{total}^L=1+R_{trace}^L+R_{ins}^L, \label{eq:r_total}
\end{align}
in which
\begin{align}
N_{\Viterbi}=&(2+1)(k+m-v)2^{v}+2\sum_{i=1}^v2^i+\sum_{i=0}^{v-1}2^i \label{eq:N_ACS}\\
 \phantom{}&+C_1\cdot\left[2(k+m+v)+1.5(k+m)\right]\\
  =&5(2^v-1)+3(k+m-v)\cdot2^{v}\notag\\
  \phantom{}&+C_1\cdot\left[2(k+m+v)+1.5(k+m)\right], \\
R_{trace}^L=&\frac{\E[N_{\LVA}]\cdot C_1\cdot\left[2(k+m+v)+1.5(k+m)\right]}{N_{\Viterbi}},\label{eq:r_trace}\\
R_{ins}^L=&\frac{\E[I_{\LVA}]\cdot C_2\cdot\log(\E[I_{\LVA}])}{N_{\Viterbi}}, \label{eq:r_ins}
\end{align}
where $C_1, C_2$ are two hardware specific constants, $\E[N_{\LVA}]$ denotes the expected number of decoding attempts and $\E[I_{\LVA}]$ denotes the expected number of insertions to maintain a sorted list. The denominator $N_{\Viterbi}$ indicates the number of operations required for a standard ACS operation.

Fig. \ref{fig: expected decoding and insertion} shows the expected number of decoding attempts versus list size $L$ and the expected number of insertions to maintain a sorted list versus list size $L$ for $(27,31)$ CC, 0x709 CRC code with $k=64$ at 2 dB. Fig. \ref{fig: complexity measurement} shows the time ratio of S-LVA as a function of list size $L$. It can be seen that \eqref{eq:r_trace} and \eqref{eq:r_ins} match the empirical time ratio of traceback operations and insertion operation with high accuracy. Though the degree of 0x709 CRC code is 10, one can observe that the overall time ratio is still comparable to that of a standard Viterbi algorithm, which indicates that using a strong CRC code may not necessarily lead to a huge complexity increase, as long as the CC-CRC pair is operated in the optimal SNR range.

\subsection{S-LVA Performance vs. $L$}\label{sec4}

As we learned in Sec. \ref{sec:L=C}, the ``complete'' S-LVA algorithm with  $L= {| {\cal C} |}$ achieves  $\Prob^{| {\cal C} |}_{\NACK}=0$ and  $\Prob^{| {\cal C} |}_{\UE}$ is well approximated by truncating the union bound of \eqref{eq:UEUB} at a reasonable $\tilde{d}$.   In the context of a feedback communication system, it is often preferable to retransmit a codeword or to lower the rate of the transmission through incremental redundancy rather than to accept undetectable errors. Thus the full complexity $L= {| {\cal C} |}$ may actually lead to detrimental results in certain cases, especially at very low SNRs where $\Prob^{| {\cal C} |}_{\UE}$ approaches 1.

Sec. \ref{sec:L=1} showed how the other extreme of $L=1$ significantly lowers the UE probability with $\Prob^{1}_{\UE}$ well approximated by the minimum between the upper bound of \eqref{eq:UE1conjecture} and the NNA of \eqref{eq:UE1_NNA}.  The reduction in $\Prob_{\UE}$ comes at the cost of a significantly increased  $\Prob^{1}_{\NACK}$, which is approximately the FER of the CC decoded by soft Viterbi without a CRC code.

We expect the best choice of $L$ for many systems to be in between these two extremes.  The rest of this section explores how  $\Prob^{L}_{\UE}$ and  $\Prob^{L}_{\NACK}$ vary with $L$.
In general, with SNR fixed, $\Prob_{\NACK}^{L}$ and $\Prob_{\UE}^{L}$ have the following properties: $\Prob_{\NACK}^{L}$ is a decreasing function of $L$ with $\lim_{L\to|\mathcal{C}|}\Prob_{\NACK}^{L}=0$, and 
$\Prob_{\UE}^{L}$ is an increasing function of $L$ with  $\lim_{L\to|\mathcal{C}|}\Prob_{\UE}^{L}=\Prob^{|\mathcal{C}|}_{\UE}$, which is well approximated by \eqref{eq:UEUB}.

Therefore, one could ask what the optimal list size $L^*$ is such that, for example,  $\Prob_{\NACK}^{L}\le\Prob_{\NACK}^*$ and $\Prob_{\UE}^{L}\le\Prob_{\UE}^*$, where $\Prob_{\NACK}^*$ and $\Prob_{\UE}^*$ are target erasure and UE probabilities, respectively.   We present useful bounds on $\Prob_{\NACK}^{L}$ and $\Prob_{\UE}^{L}$ to further explore the concept of an optimal list size $L^*$.

\begin{corollary}[Markov bound on $\Prob_{\NACK}^{L}$]
The erasure probability $\Prob_{\NACK}^{L}$ satisfies $\Prob_{\NACK}^{L}\le\frac1{L}$ if $\gamma_s\to\infty$. 
\end{corollary}
\begin{IEEEproof}
The result is a direct consequence of Markov inequality. The erasure probability with a list size $L$ is given as $\Prob_{\NACK}^{L}=\Pr\{N_{\LVA}>L\}$, where $N_{\LVA}$ is the random variable representing the decoding trial at which the CRC check first passes. By applying Markov inequality for $\gamma_s\to\infty$, we have
\begin{align}
\Prob_{\NACK}^{L}=\Pr\{N_{\LVA}>L\}\le\frac{\E[N_{\LVA}]}{L}=\frac1{L}.
\end{align}
\end{IEEEproof}

\begin{figure}[t]
\centering
\includegraphics[scale=0.33]{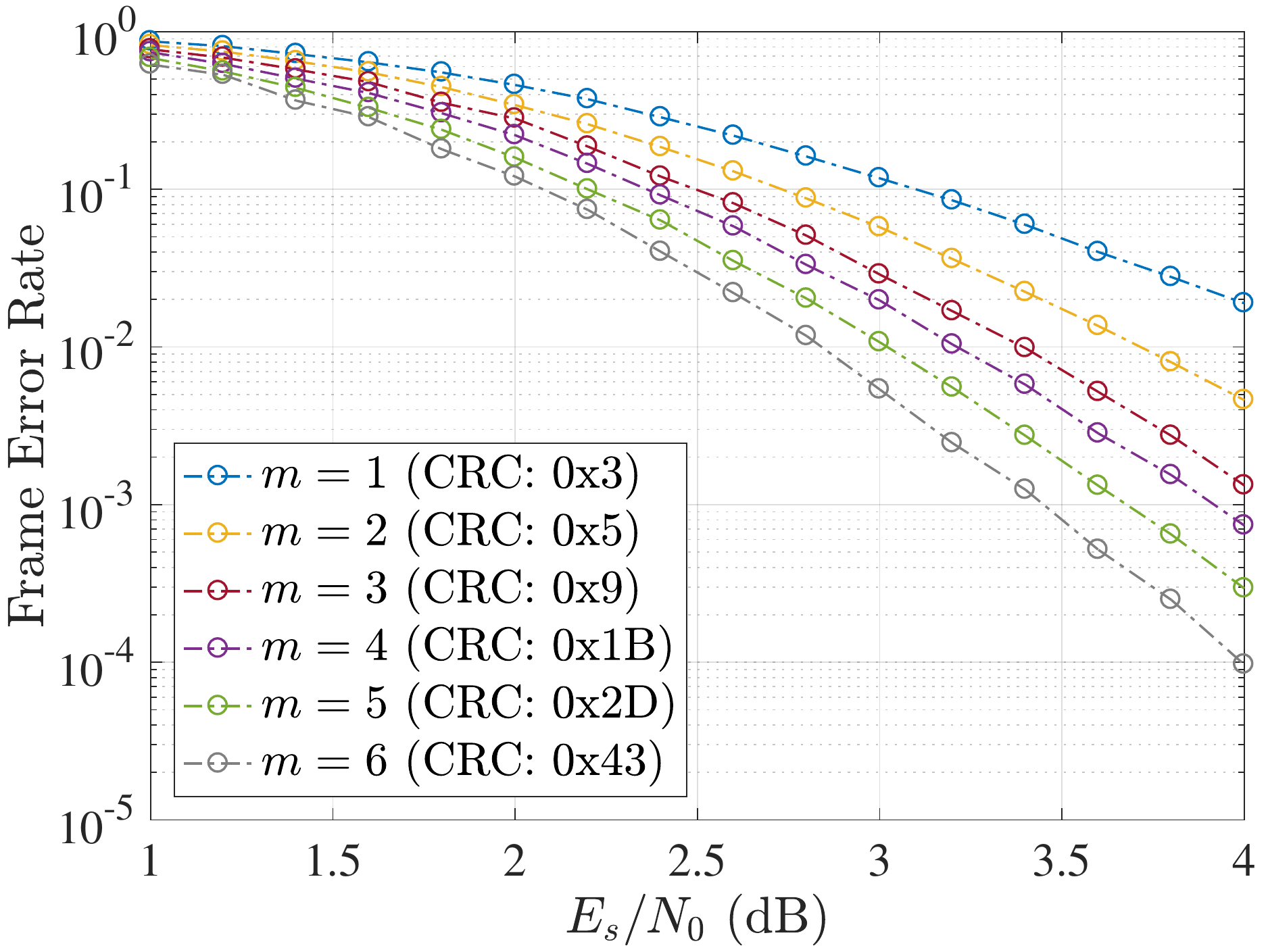}
\caption{FER vs. $E_s/N_0$ of degree $1-6$ optimal CRC codes for $(13, 17)$ CC with $k=256$.}
\label{fig5}
\end{figure}

\begin{figure}[t]
\centering
\includegraphics[scale=0.33]{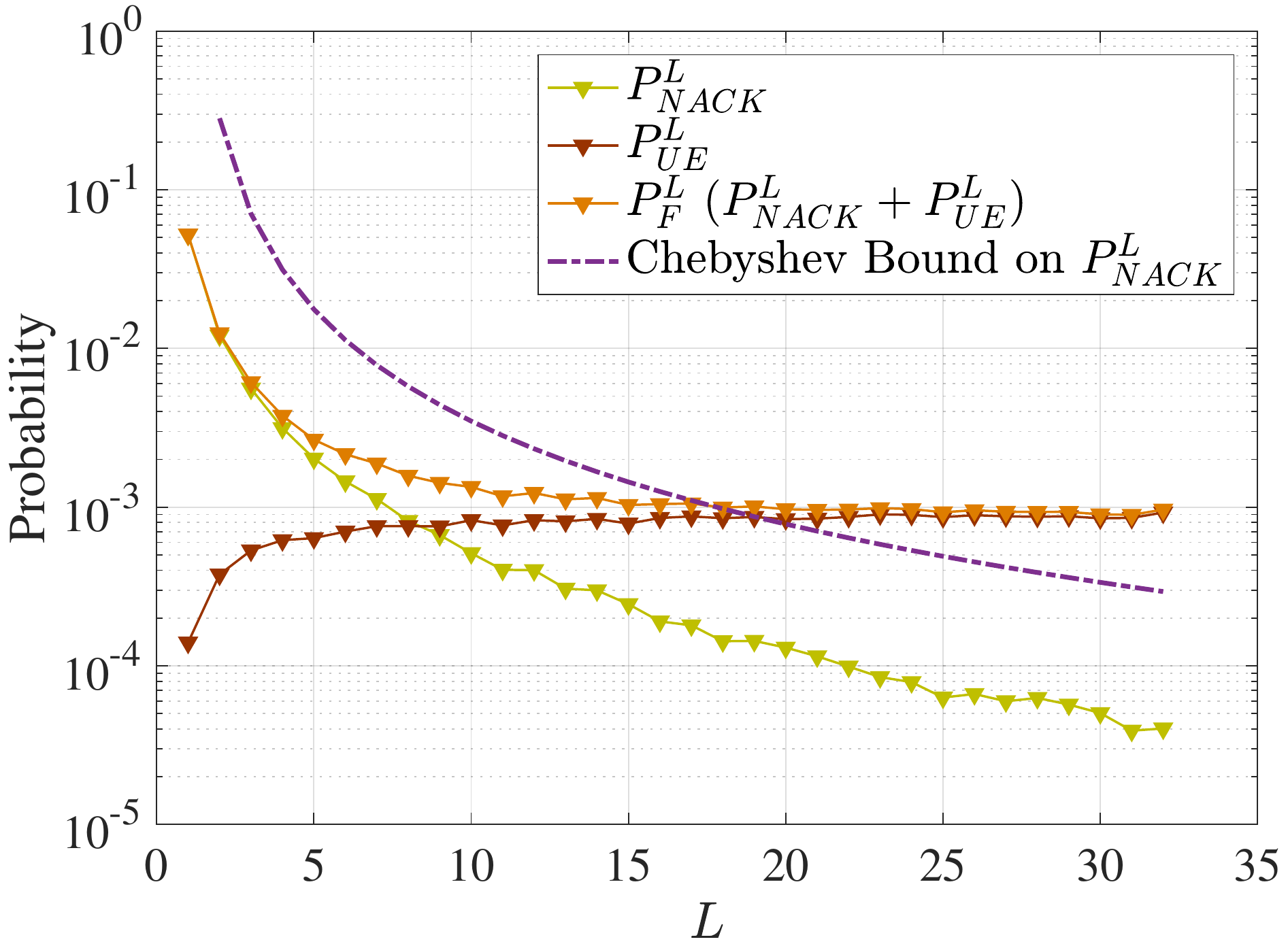}
\caption{Trade-off between $\Prob_{\NACK}^{L}$ and $\Prob_{\UE}^{L}$ for the optimal degree-$5$ CRC code 0x2D and $(13, 17)$ CC when $k=256,\ \gamma_s=3.7$ dB.
}
\label{fig6}
\end{figure}

A more useful Chebyshev bound on $\Prob_{\NACK}^{L}$ could be obtained if one knows the variance $\var(N_{\LVA})$ at high SNR.
\begin{corollary}[Chebyshev bound on $\Prob_{\NACK}^{L}$]
Given $\var(N_{\LVA})$ at $\gamma_s\gg0$, $\Prob_{\NACK}^{L}$ satisfies $\Prob_{\NACK}^{L}\le\frac{\var(N_{\LVA})}{(L-1)^2}$, where $L\ge2$.
\end{corollary}

\begin{IEEEproof}
The result is a direct consequence of Chebyshev inequality. Since $\gamma_s\gg0$, $\E[N_{\LVA}]\to1$. From Chebyshev inequality, we have
\begin{align}
\Prob_{\NACK}^{L}&=\Pr\{N_{\LVA}> L\}\notag\\
&=\Pr\{N_{\LVA}\ge L+1\}\notag\\
&\le\Pr\{|N_{\LVA}-\E[N_{\LVA}]|\ge L-\E[N_{\LVA}]+1\}\notag\\
&\le\frac{\var(N_{\LVA})}{(L-(\E[N_{\LVA}]-1))^2}\notag\\
&\le\frac{\var(N_{\LVA})}{(L-1)^2}.
\end{align}
\end{IEEEproof}

As an example, we study the trade-off between $\Prob_{\NACK}^{L}$ and $\Prob_{\UE}^{L}$ for the $(13,17)$ CC. Assume at $\gamma_s=3.7$ dB, $\Prob_{\NACK}^*=10^{-3}$ and $\Prob_{\UE}^*=8\times10^{-4}$. In Fig.~\ref{fig5}, the FER of degree $1-6$ optimal CRC codes is plotted. Here we use the optimal degree-$5$ CRC code with the $(13,17)$ CC to illustrate how to find the optimal list size $L^*$. Fig.~\ref{fig6} shows the trade-off between $\Prob_{\NACK}^{L}$ and $\Prob_{\UE}^{L}$ when $k=256$ at $3.7$ dB. It can be seen that $L^*=8$ satisfies $\Prob_{\NACK}^{L}\le\Prob_{\NACK}^*$ and $\Prob_{\UE}^{L}\le\Prob_{\UE}^*$. 

If $\Prob_{\NACK}^*=10^{-3}$, $\Prob_{\UE}^*=10^{-3}$ and empirical $\var(N_{\LVA})=0.2823$ is known, since $\Prob_{\UE}^{L}\le\Prob_{\UE}^*$ always holds, one can directly apply the empirical Chebyshev bound to obtain $L^*\ge18$ without knowing the true $\Prob_{\NACK}^{L}$ curve.

\section{Coded Channel and Its Capacity}\label{sec:coded channel}

In Sec. \ref{sec:S-LVA}, we have thoroughly discussed the performance of S-LVA combined with the optimal CRC code designed specifically for the given CC, in which the decoding complexity depends mainly on the expected number of decoding attempts. One important observation is that, with SNR in a relatively high regime, this expected number is much less than $2^m(1-\epsilon)$, where $\epsilon>0$ is a small constant, which suggests that the decoding can be done much more efficiently. Still, different CC-CRC pair corresponds to different decoding compleixty. Therefore, a more general question to ask is that, how to select the optimal CC-CRC pair for the system model introduced in Sec. \ref{sec:system model}. We propose the coded channel model to address this problem.

\begin{figure}[t]
\centering
\includegraphics[scale=0.6]{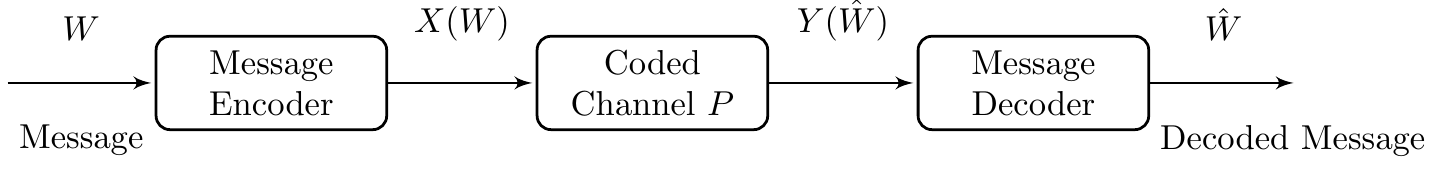}
\caption{Block diagram of the equivalent coded channel}
\label{fig:coded channel model}
\end{figure}

\subsection{The Coded Channel Model}
The equivalent coded channel model of the system model introduced in Sec. \ref{sec:system model} is shown in Fig. \ref{fig:coded channel model}, which consists of two finite sets $\mathcal{X}$ and $\mathcal{Y}$ and a channel matrix $P$, where $\mathcal{X}$ denotes the set of all possible $k$-bit message sequences with $|\mathcal{X}|=2^k$, $\mathcal{Y}=\mathcal{X}\cup\{E\}$ with $|\mathcal{Y}|=2^k+1$ and the channel matrix $P$ is a single equivalent abstraction of the CRC encoder, the convolutional encoder, the AWGN channel, the S-LVA decoder and the CRC decoder in Fig. \ref{fig:system model}. To make the coded channel complete, we introduce the ``outer'' message encoder which simply selects the $W$-th message symbol $X(W)$ in $\mathcal{X}$ and the ``outer'' message decoder which simply decodes message symbol $Y(\hat{W})$ to the $\hat{W}$-th message, where $W\in\{1,2,\cdots,2^k\}$ and $\hat{W}\in\{1,2,\cdots,2^k,2^k+1\}$ are both indices. If $W=\hat{W}$, then $X(W)=Y(\hat{W})$ and vice versa. If $Y(\hat{W})=E$, then $\hat{W}=2^k+1$.

Obviously, if one knows each transition probability from $X^k$ to $Y^k$ and $X^k$ to $E$, then the entire part from the CRC encoder to CRC decoder shown in Fig. \ref{fig:system model} can be equivalently substituted with a single channel $P$ and the corresponding coded channel capacity $C(P)$, which indicates the maximum bits per codeword transmission, can be computed. 

For brevity, define $\epsilon\triangleq P^L_{\UE}$ and $\alpha\triangleq P^L_{\NACK}$ which indicate the overall characteristics of the coded channel $P$. Unless otherwise stated, we will keep this notation in the following sections. We first show that $P$ is a symmetric channel.

\begin{theorem}
The equivalent coded channel matrix $P$ of the CRC encoder, the convolutional encoder, the AWGN channel, the S-LVA decoder, and the CRC decoder, is a symmetric channel, and the coded channel capacity $C(P)$ is achieved by the uniform distribution.
\end{theorem}
\begin{IEEEproof}
Let us partition $P$ into $P=[Q\mid\alpha I]$ where $\alpha\triangleq\Prob^L_{\NACK}$, $Q$ denotes a $2^k\times 2^k$ matrix, and $I$ is a $2^k\times 1$ all-one matrix. It can be shown that $P$ satisfies the following properties: 
\begin{itemize}
\item[(i)] $Q=Q^T$ due to the linearity of the convolutional code; 
\item[(ii)] Rows in $Q$ are permutations of each other, which is due to the independence of the Gaussian noise on the transmitted codeword; 
\item[(iii)] Columns in $Q$ are permutations of each other, which is a direct consequence of (i) and (ii). 
\end{itemize}
Since $\alpha I$ also satisfies (ii) and (iii). Therefore $P=[Q\mid\alpha I]$ is a symmetric channel and the capacity is achieved by the uniform distribution.
\end{IEEEproof}

\begin{figure}[t]
\centering
\includegraphics[scale=0.35]{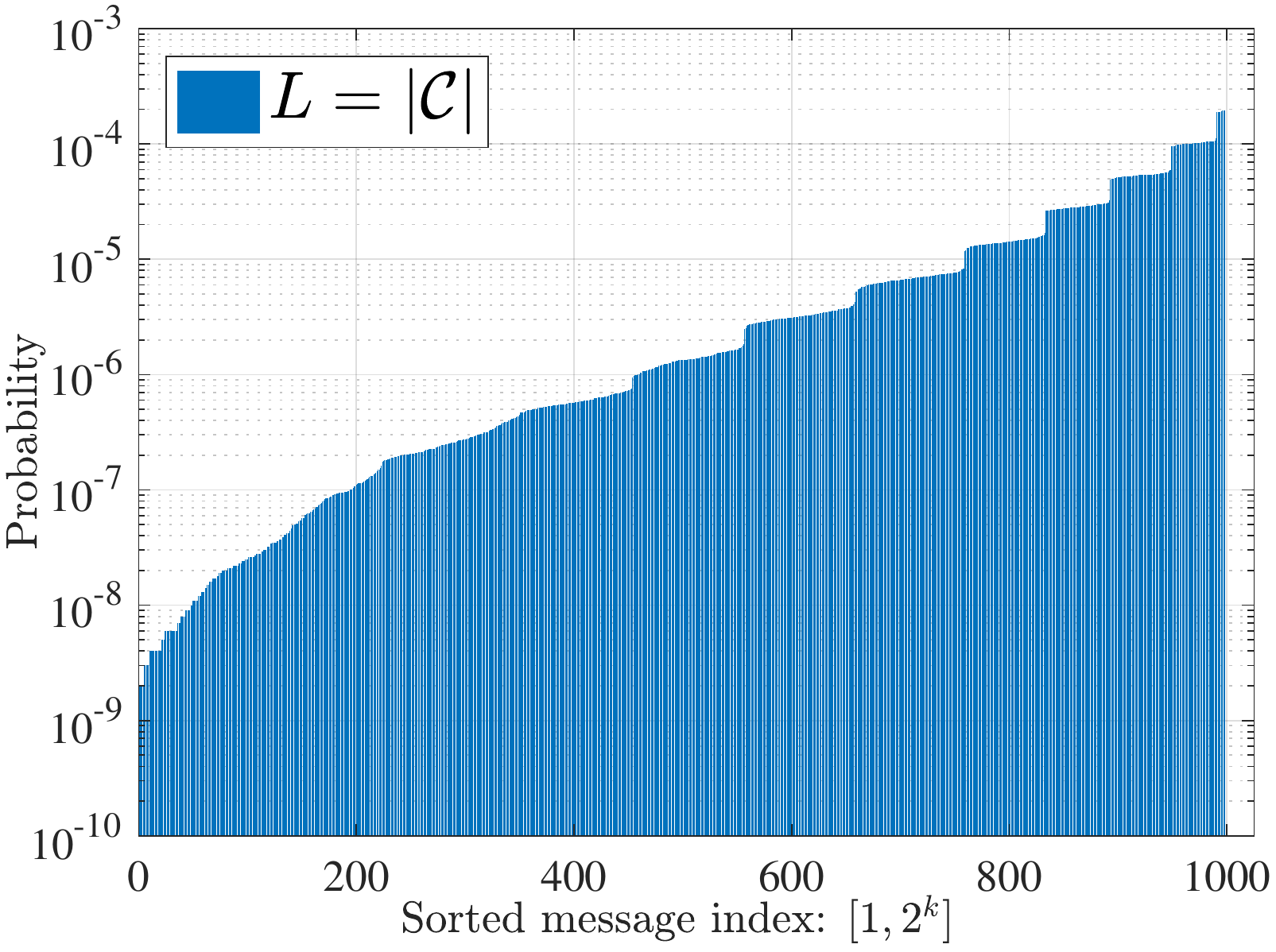}
\caption{A typical stair-shaped of probability distribution of the unknown probabilities with $(13, 17)$ CC, a degree-$6$ CRC code 0x43, $k=10$ and S-LVA with $L=|\mathcal{C}|$ at $0$ dB when transmitted convolutional code is the all-zero codeword. Some zero unknown probabilities are omitted due to the insufficient number of simulations.The highest level corresponds to the probability of decoding to nearest neighbors of the transmitted all-zero convolutional code.}
\label{fig: prob distribution}
\end{figure}

\subsection{True Coded Channel}
In practice, it is difficult to completely determine each entry of $P$, especially when $k$ is large. Therefore let the unknown probabilities be specified as $p_1, p_2,\cdots,p_{2^k-1}$ with $p_i\ge0$ and $\sum_{i=1}^{2^k-1}p_i=\epsilon$, for each transmitted message. Thus, the true coded channel capacity $C(P)$ can be computed when $p(x)$ is uniformly distributed
\begin{align}
C(P)=&H(Y)-H(Y|X=x(w))\\
=&H_{2^k+1}\left(\frac{1-\alpha}{2^k},\cdots,\frac{1-\alpha}{2^k},\alpha\right)\notag\\
\phantom{}&-H_{2^k+1}(1-\epsilon-\alpha, \alpha,p_1,p_2,\cdots,p_{2^k-1})\\
=&(1-\alpha)\left[k-H\left(\frac{\epsilon}{1-\alpha}\right)\right]\notag\\
\phantom{}&-\epsilon H_{2^k-1}\left(\frac{p_1}{\epsilon},\frac{p_2}{\epsilon},\cdots,\frac{p_{2^k-1}}{\epsilon}\right),
\end{align}
where $x(w)$ is some fixed message symbol in $\mathcal{X}$.

Now that the true coded channel is a much complicated model, still, there are some intuitions that can be drawn from this model. As an example, Fig. \ref{fig: prob distribution} shows the sorted probability distribution of the unknown probabilities $p_1,p_2,\cdots,p_{2^k-1}$ for $k=10$, which demonstrates a stair-shaped envelop. The highest level corresponds to the probabilities of decoding to the nearest neighbors of the transmitted convolutonal codeword. As SNR increases, the bulk of probability of error will move towards nearest neighbors, which suggests that nearest neighbors might be a useful tool to approximate the true coded channel capacity.

To formally present the above intuitions, we propose the following three simplied coded channel models which only require the knowlege of $\epsilon$, $\alpha$ and the number of nearest neighbors of the transmitted message $N$ to approximate the true coded channel, which are referred to as loose lower bound model (LLB), nearest neighbor lower bound model (NNLB) and nearest neighbor upper bound model (NNUB).

\subsection{Loose Lower Bound Model (LLB)}
In this model, we assume that for each transmitted message symbol, the probability of decoding to the erasure symbol $E$ is $\alpha$ and the probabilities of decoding to message symbols other than the transmitted message are equally likely with $p_i=\frac{\epsilon}{2^k-1}$ for $i=1,2,\cdots,2^k-1$. 

% Namely, the LLB coded channel $P_{\LLB}$ is
% \begin{align}
% P_{\LLB}\triangleq\begin{bmatrix}
% 1-\epsilon-\alpha & \frac{\epsilon}{2^k-1} & \cdots & \frac{\epsilon}{2^k-1} & \alpha \\
% \frac{\epsilon}{2^k-1} & 1-\epsilon-\alpha & \cdots & \frac{\epsilon}{2^k-1} & \alpha \\
% \frac{\epsilon}{2^k-1} & \frac{\epsilon}{2^k-1} & \cdots & \frac{\epsilon}{2^k-1}& \alpha \\
% \vdots  & \vdots &\ddots & \vdots & \vdots\\
% \frac{\epsilon}{2^k-1} & \frac{\epsilon}{2^k-1} & \cdots & 1-\epsilon-\alpha & \alpha
% \end{bmatrix}.
% \end{align}

Similarly, the capacity $C(P_{\LLB})$ can be computed as
\begin{align}
C(P_{\LLB})=(1-\alpha)\left[k-H\left(\frac{\epsilon}{1-\alpha}\right)\right]-\epsilon\log(2^k-1).
\end{align}

Obviously, $C(P_{\LLB})<C(P)$. The reason why this model becomes loose is that, except for the probability of decoding correctly or decoding to an erasure symbol, the rest of the probability is evenly allocated to message symbols other than the transmitted one. However in the true coded channel model, the nearest neighbors of the transmitted convolutional code will account for most of the rest probability since they are the closest codewords that S-LVA decodes to.

\begin{figure}[t]
\centering
\includegraphics[scale=0.45]{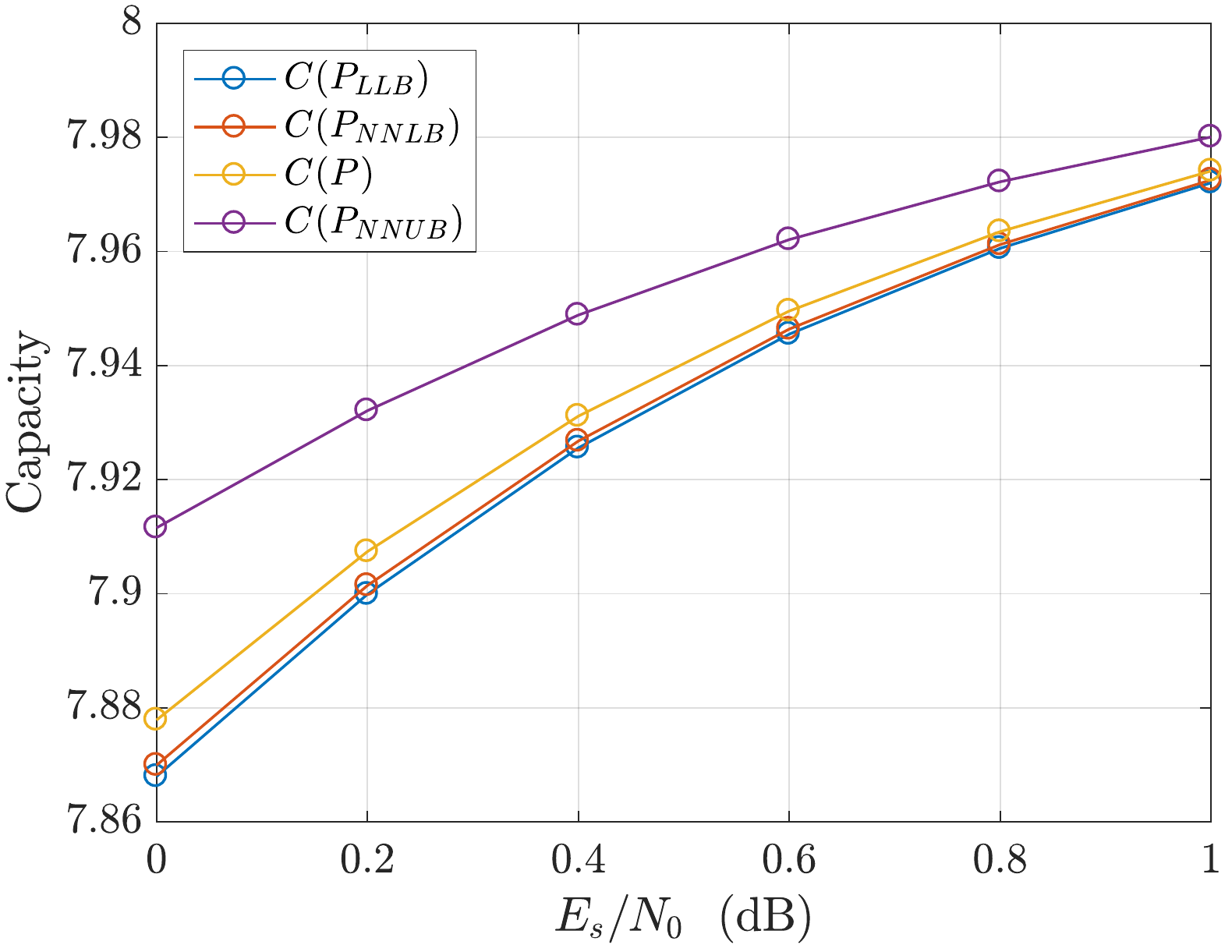}
\caption{Capacity vs. SNR for LLB, NNLB, true coded channel, and NNUB model, with $k=8$, $m=6$, and $v=3$, in which $n_c=2(k+m+v)$ denotes the number of bits that are sent to the binary AWGN (Bi-AWGN) channel. }
\label{fig:capacity vs. SNR}
\end{figure}

\subsection{Nearest Neighbor Lower Bound Model (NNLB)}
In this model, we assume that for each transmitted message symbol, the number of nearest neighbors $N\ (0<N<2^k-1)$ and the approximate probability of a single nearest neighbor $\epsilon^*$ are known. Here, $\frac{\epsilon}{2^k-1}<\epsilon^*<\frac{\epsilon}{N}$ since the nearest neighbors have the highest probability thus $\epsilon^*$ should be above the average. Thus, the remaining $2^k-1-N$ unknown probabilities will equally split probability $\epsilon-N\epsilon^*$. The capacity for this channel, $C(P_{\NNLB})$, can be computed as
\begin{align}
C(P_{\NNLB})=&(1-\alpha)\left[k-H\left(\frac{\epsilon}{1-\alpha}\right)\right]-\epsilon H\left(\frac{N\epsilon^*}{\epsilon}\right)\notag\\
\phantom{}&-N\epsilon^*\log N-(\epsilon-N\epsilon^*)\log(2^k-1-N).
\end{align}

We point out that the NNLB model will awlays be a good approximation on the true coded channel capacity, since the nearest neighbors are taken into account which have the dominating unknown probabilities. As SNR increases, the nearest neighbors will be the most likely erroneously decoded codewords and codewords further away than nearest neighbors will be more unlikely. Therefore, we expect $C(P_{\NNLB})$ to approach $C(P)$ in high SNR regime. In fact, an extreme situation would be that $\epsilon$ only goes to the nearest neighbors, which gives rise to the following upper bound model.

\subsection{Nearest Neighbor Upper Bound Model (NNUB)}
In this model, we assume that for each transmitted message symbol, the number of nearest neighbors $N$ is known and probability of error $\epsilon$ is equally divided only by the nearest neighbors. That is, probability of each nearest neighbor is $\frac{\epsilon}{N}$ and codewords further away from nearest neighbors are unlikely. Thus, the capacity for this channel, $C(P_{\NNUB})$, can be computed as
\begin{align}
C(P_{\NNUB})=&(1-\alpha)\left[k-H\left(\frac{\epsilon}{1-\alpha}\right)\right]-\epsilon\log N.
\end{align}

\subsection{Comparisons}
The following theorem describes the relationships among the above four models. 
\begin{theorem}
For a coded channel with message blocklength $k$, it holds that
\begin{align}
C(P_{\LLB})<C(P_{\NNLB})<C(P)<C(P_{\NNUB})+\epsilon\log N,
\end{align}
provided that the $2^k-1$ unknown probabilities of each row in coded channel $P$ are distinct, $0<N<2^k-1$, and $\frac{\epsilon}{2^k-1}<\epsilon^*<\frac{\epsilon}{N}$.
\end{theorem}
\begin{IEEEproof}
The chain of inequalities $C(P_{\LLB})<C(P_{\NNLB})<C(P)$ can be established by applying the fact that the uniform increases entropy to $H(Y|X=x(w))$.
\end{IEEEproof}

As an example, Fig. \ref{fig:capacity vs. SNR} illustrates the capacities for LLB channel, NNLB channel, true coded channel, and NNUB channel.

\section{Optimal CC-CRC Design}\label{sec:optimal CC-CRC}

\renewcommand\arraystretch{1.1}
\begin{table}[t]
\caption{Most Popular Rate-$1/2$ Convolutional Codes and Corresponding Distance-Spectrum-Optimal CRC Codes with $k=64$}
\scalebox{0.72}{
\begin{tabular}{r|c|c|cccccccc}
\hline
\multirow{2}{*}{$v$} & \multirow{2}{*}{Conv. Code} &\multicolumn{9}{c}{Distance-Spectrum-Optimal CRC Generator Polynomial} \\
\cline{3-11}
 & & $m$ & 3 & 4 & 5 & 6 & 7 & 8 & 9 & 10\\\hline\hline
3 & \multicolumn{1}{l}{(13,17)} & & 0x9 & 0x1B & 0x2D & 0x43 & 0xB5 & 0x107 & 0x313 & 0x50B\\
4 & \multicolumn{1}{l}{(27,31)} & & 0xF & 0x15 & 0x33 & 0x4F & 0xD3 & 0x13F & 0x2AD & 0x709\\
5 & \multicolumn{1}{l}{(53,75)} & & 0x9 & 0x11 & 0x25 & 0x49 & 0xEF & 0x131 & 0x23F & 0x73D \\
6 & \multicolumn{1}{l}{(133,171)} & & 0xF & 0x1B & 0x23 & 0x41 & 0x8F & 0x113 & 0x2EF & 0x629\\
7 & \multicolumn{1}{l}{(247,371)} & & 0x9 & 0x13 & 0x3F & 0x5B & 0xE9 & 0x17F & 0x2A5 & 0x61D\\
8 & \multicolumn{1}{l}{(561,753)} & & 0xF & 0x11 & 0x33 & 0x49 & 0x8B & 0x19D & 0x27B & 0x4CF\\
9 & \multicolumn{1}{l}{(1131,1537)} & & 0xD & 0x15 & 0x21 & 0x51 & 0xB7 & 0x1D5 & 0x20F & 0x50D\\
10 & \multicolumn{1}{l}{(2473,3217)} & & 0xF & 0x13 & 0x3D & 0x5B & 0xBB & 0x105 & 0x20D & 0x6BB\\
\hline
\end{tabular}}
\label{table: CC-CRC pair}
\end{table}

\begin{figure}[t]
\centering
\includegraphics[scale=0.41]{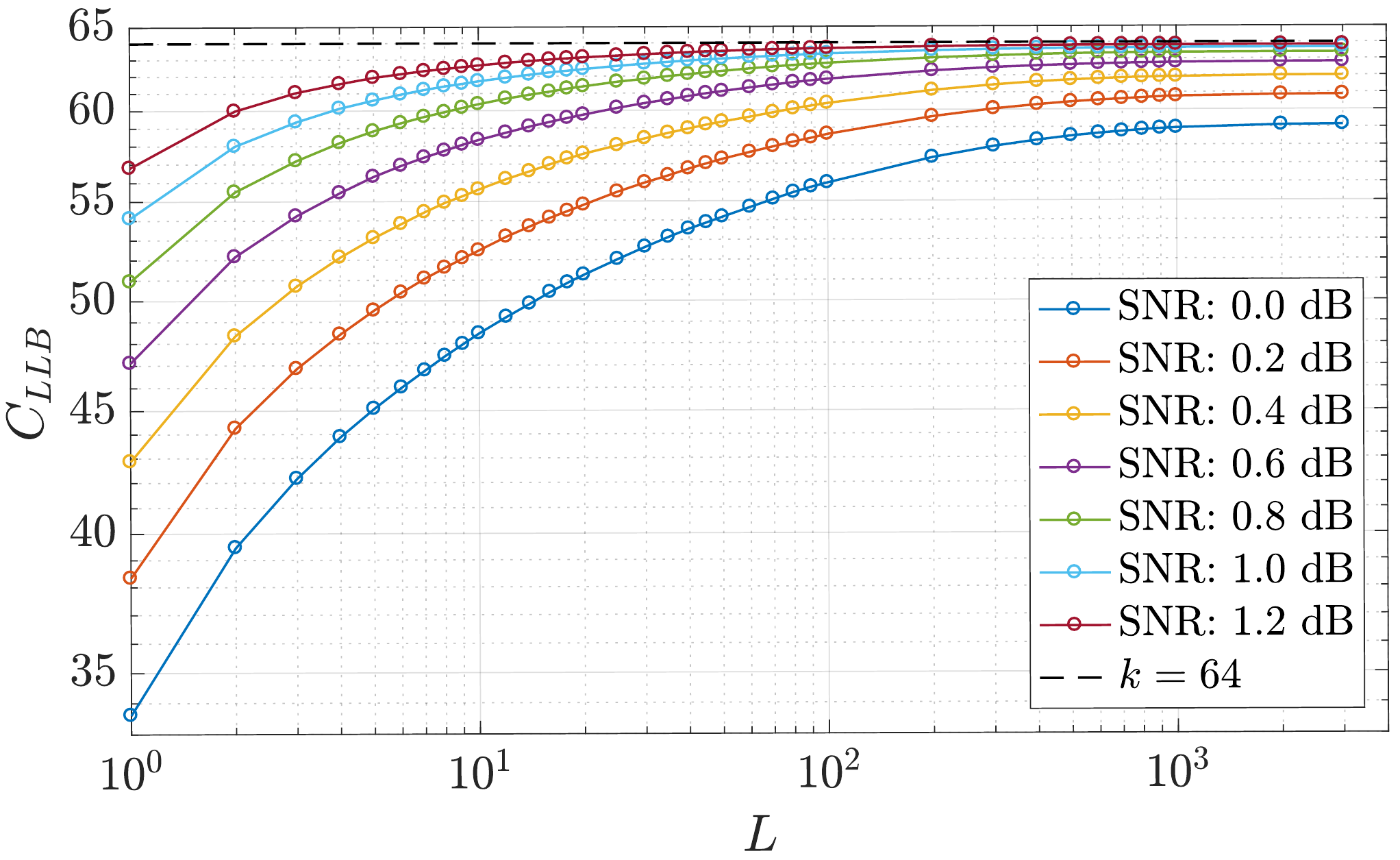}
\caption{The coded channel capacity $C_{\LLB}$ in loose lower bound model vs. list size $L$ for $(247, 371)$ CC and 0x61D CRC code.}
\label{fig:capacity vs. complexity}
\end{figure}

\begin{figure}[t]
\centering
\includegraphics[scale=0.4]{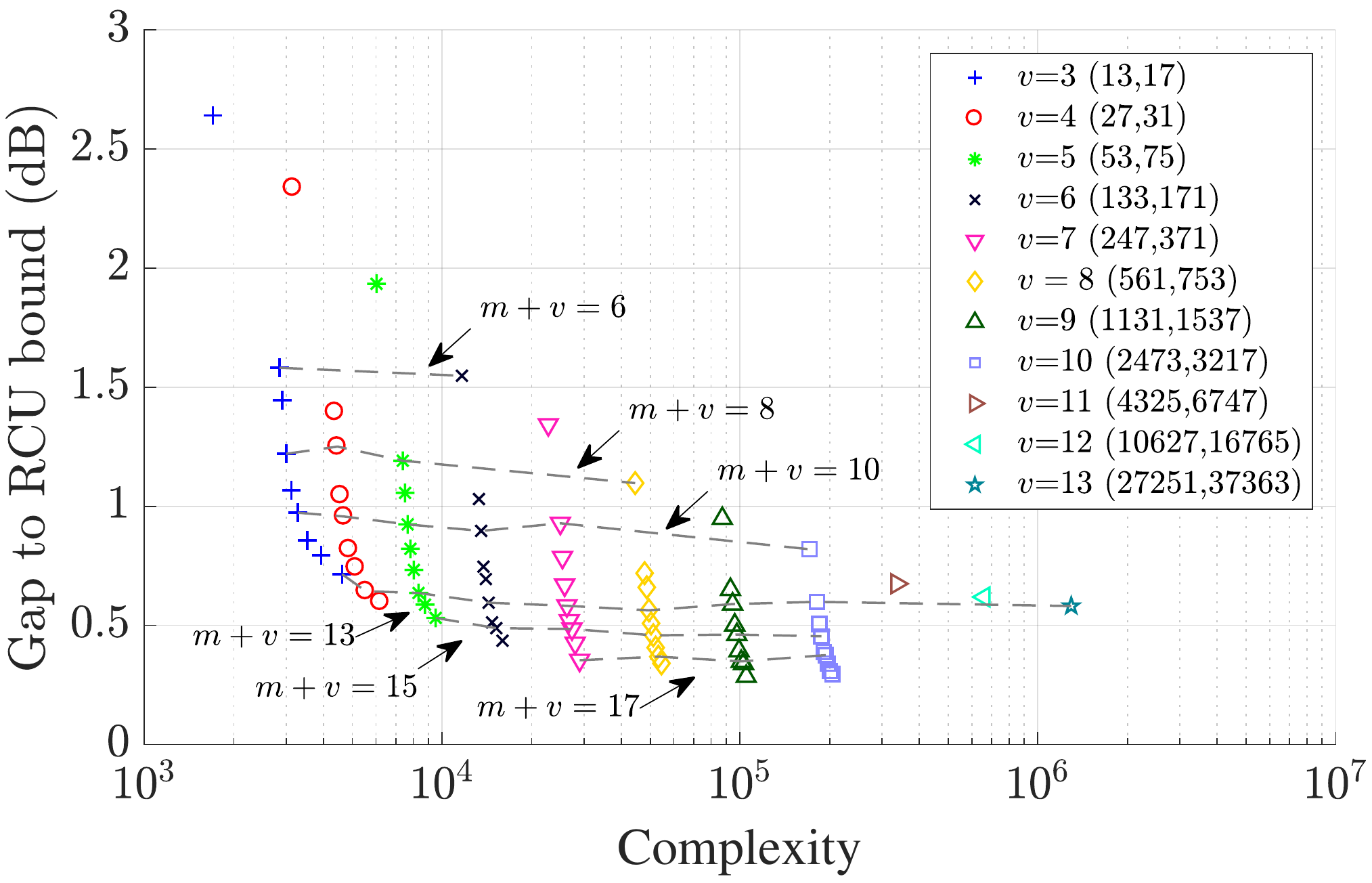}
\caption{the SNR ($E_s/N_0$) gap to RCU bound vs. decoding complexity for various CC-CRC pairs with $k=64$ and target FER $10^{-3}$. Each color corresponds to a specific CC shown in parenthesis. Markers from top to bottom with the same color correspond to soft Viterbi decoding, $m=3,4,\cdots,10$ distance-spectrum-optimal CRC codes, respectively. CCs with $v=11,12,13$ using soft Viterbi decoding are also provided.}
\label{fig:gap vs. complexity}
\end{figure}

In this section, we present the design methodology and examples of optimal CC-CRC pairs under a target FER. Since the design of optimal list size $L$ is independent of the design of optimal CC-CRC pairs, we first show that $L=|\mathcal{C}|$ is always the optimal list sizes for any CC-CRC pairs regardless of SNR by using the coded channel capacity argument. Then, given that $L=|\mathcal{C}|$ where FER is simply probability of error, we choose the design metric as the SNR gap to RCU bound derived by Polyanskiy \emph{et al.} in \cite{Polyanskiy2010} and well-approximated by the saddlepoint method in \cite{Segura2018} when the target FER is achieved. The optimal CC-CRC pair is the one that has the smallest SNR gap with the least complexity. The convolutional codes considered in this paper are from \cite{ErrorControlCoding}. 

Table \ref{table: CC-CRC pair} presents the candidate rate-$1/2$ convolutional codes with $v$ ranging from $3$ to $10$, each with the distance-spectrum-optimal CRC codes with degree $m$ ranging from $3$ to $10$ using Lou \emph{et al.}'s method for $k=64$.

First, for any CC-CRC pairs, the best performance is always achieved with $L=|\mathcal{C}|$, regardless of SNR. Fig. \ref{fig:capacity vs. complexity} illustrates the coded channel capacity $C_{\LLB}$ in loose lower bound model versus list size $L$ for $(247,371)$ CC and 0x61D CRC code. Under various SNR values, $C_{\LLB}$ grows monotonically with $L$, which indicates that $L=|\mathcal{C}|$ is the optimal list size. Note that although $L$ reaches the maximum value, the decoding complexity only depends on the $\E[N_{\LVA}]$ and $E[I_{\LVA}]$ and they both converge when $L$ is large enough.

With $L=|\mathcal{C}|$ fixed, the design metric could be the SNR gap to the RCU bound and the optimal CC-CRC pair should be the one that minimizes this gap with the least complexity. In most cases, it is difficult to take care of SNR gap and complexity simultaneously. Thus, one alternative is to set a target SNR gap and the optimal CC-CRC pair is the one that is less than the target SNR gap with the minimum complexity.

Fig. \ref{fig:gap vs. complexity} demonstrates that with target FER of $10^{-3}$ fixed, the SNR ($E_s/N_0$) gap to RCU bound versus decoding complexity for various CC-CRC pairs presented in Table \ref{table: CC-CRC pair}. In the plot, the decoding complexity is measured by the scaled number of operations, which is equal to $R^{|\mathcal{C}|}_{tot}\cdot N_{\Viterbi}$ with $N_{\Viterbi}$ defined in \eqref{eq:N_ACS}. Setting 0.5 dB as the target SNR gap, we noticed that CC-CRC pairs that are less than 0.5 dB away from RCU bound are $(v=6, m\ge9), (v=7, m\ge 8), (v=8, m\ge 7), (v=9, m\ge 6), (v=10, m\ge 5)$, among which $(v=6, m=9)$ has the minimum complexity.Therefore in this example the best CC-CRC pair is $(v=4, m=9)$ in Table \ref{table: CC-CRC pair}. 

Besides, Fig. \ref{fig:gap vs. complexity} also shows that CC-CRC pairs with the same $m+v$ have nearly the same SNR gap which indicates that they have roughly the same performance and only complexity differs. Therefore, we propose the following conjecture regarding the performance of CC-CRC pairs with constant $m+v$, i.e., constant number of redundant bits.

\begin{conjecture}\label{conjecture: m+v the same}
Any minimal convolutional code of $m$ memory elements used with the degree-$v$ distance-spectrum-optimal CRC code under serial list Viterbi decoding operated at the same SNR will have the same FER performance, provided that $m+v$ is the same.
\end{conjecture}

If Conjecture \ref{conjecture: m+v the same} is corroborated, since decoding complexity grows exponentially with $v$. Then the optimal CC-CRC pair with the minimum decoding complexity is a weaker CC used with a large degree distance-spectrum-optimal CRC code. 

\begin{figure}[t]
\centering
\includegraphics[scale=0.38]{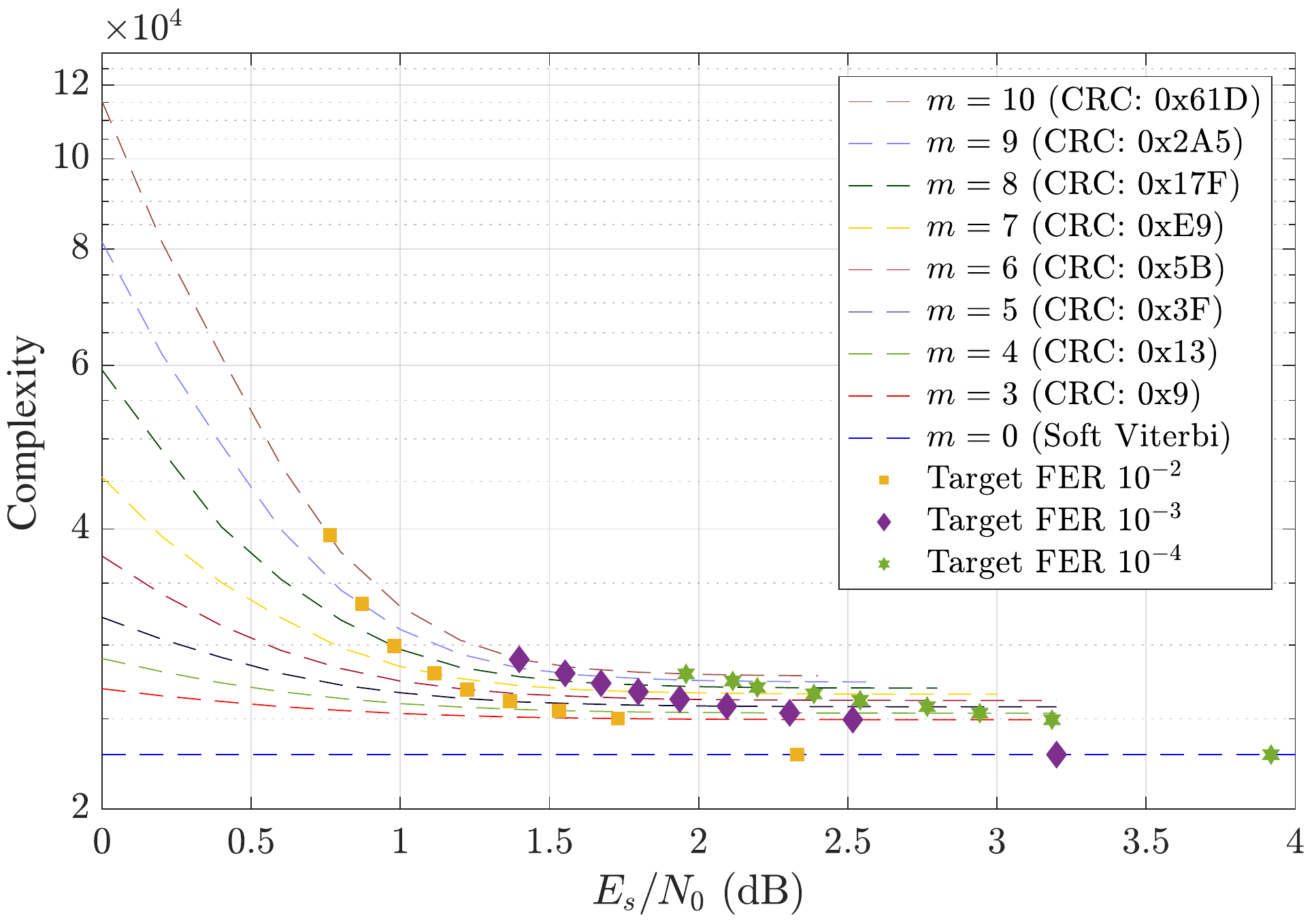}
\caption{The decoding complexity vs. SNR for $(247,371)$ CC with its correspondinng distance-spectrum-optimal CRC codes. The CC with no CRC using soft Viterbi decoding is also given as a reference.}
\label{fig:complexity vs. SNR}
\end{figure}

Although Fig. \ref{fig:gap vs. complexity} demonstrates the SNR gap to RCU bound for each CC-CRC pair to reach the target FER $10^{-3}$. Still, one may wonder whether the actual SNR that achieves the target FER for some CC-CRC pair could be impractically high. Let $\gamma_s^*$ be the SNR that achieves the target FER for a CC-CRC pair. Fig. \ref{fig:complexity vs. SNR} provides an empirial answer to this question. In Fig. \ref{fig:complexity vs. SNR}, the decoding complexity for $(247, 371)$ CC used with its corresponding distance-spectrum-optimal CRC codes is plotted and the actual SNR points for each CC-CRC pair to reach target FER $10^{-2}$, $10^{-3}$ and $10^{-4}$ are highlighted. We can observe that: (i) convolutional codes used with a distance-spectrum-optimal CRC code can reduce $\gamma_s^*$ considerably at the expense of a reasonable complexity; (ii) if target FER decreases one order of magnitude, the SNR increase for CC used with a distance-spectrum-optimal CRC code is smaller than that for CC with no CRC code using soft Viterbi decoding.

\section{Conclusion}\label{sec:conclusion}
For a convolutionally encoded system with CRC using serial list Viterbi decoding, an optimal CC-CRC pair and the optimal list size $L$ of S-LVA should maximize the coded channel capacity of the system. 

We first analyze the performance of S-LVA in great detail and prove that the expected number of decoding attempts, $\E[N_{\LVA}]$ converges to $2^m(1-\epsilon)$ as SNR decreases and to $1$ as SNR increases. Then we show that with SNR fixed, probability of error converges and probability of erasure tends to zero as $L$ increases up to $|\mathcal{C}|$.

Since the design of list size $L$ is independent of the design of the optimal CC-CRC pair, we deal with two design problems seperately. We first show that $L=|\mathcal{C}|$ is always the optimal list size for any candidate CC-CRC pairs. Then, with $L=|\mathcal{C}|$, since when FER is small, the corresponding coded channel capacity will be roughly the same for all candidate CC-CRC pairs, we choose the design metric of finding the optimal CC-CRC pair as the SNR gap to RCU bound proposed by Polyanskiy \emph{et al.} and provides sufficient evidences showing that a weaker CC used with a stronger distance-spectrum-optimal CRC code is comparable to a single strong CC with no CRC code.

Future work will be focused on resolving the variable rate issue by considering tail-biting CC or punctured CC.

\section*{Acknowledgment}

The authors would like to thank Fabian Steiner for pointing to us \cite{Segura2018} for efficiently approximating the random coding union (RCU) bound proposed in \cite{Polyanskiy2010}.

% Can use something like this to put references on a page
% by themselves when using endfloat and the captionsoff option.
\ifCLASSOPTIONcaptionsoff
  \newpage
\fi

\bibliographystyle{IEEEtran}
\bibliography{IEEEabrv,references}

% Generated by IEEEtran.bst, version: 1.14 (2015/08/26)
\begin{thebibliography}{10}
\providecommand{\url}[1]{#1}
\csname url@samestyle\endcsname
\providecommand{\newblock}{\relax}
\providecommand{\bibinfo}[2]{#2}
\providecommand{\BIBentrySTDinterwordspacing}{\spaceskip=0pt\relax}
\providecommand{\BIBentryALTinterwordstretchfactor}{4}
\providecommand{\BIBentryALTinterwordspacing}{\spaceskip=\fontdimen2\font plus
\BIBentryALTinterwordstretchfactor\fontdimen3\font minus
  \fontdimen4\font\relax}
\providecommand{\BIBforeignlanguage}[2]{{%
\expandafter\ifx\csname l@#1\endcsname\relax
\typeout{** WARNING: IEEEtran.bst: No hyphenation pattern has been}%
\typeout{** loaded for the language `#1'. Using the pattern for}%
\typeout{** the default language instead.}%
\else
\language=\csname l@#1\endcsname
\fi
#2}}
\providecommand{\BIBdecl}{\relax}
\BIBdecl

\bibitem{Yang2018}
H.~Yang, S.~V.~S. Ranganathan, and R.~D. Wesel, ``Serial list viterbi decoding
  with crc: Managing errors, erasures, and complexity,'' in \emph{GLOBECOM 2018
  - 2018 IEEE Global Communications Conference}, Dec. 2018.

\bibitem{Blahut2003}
R.~E. Blahut, \emph{Algebraic Codes for Data Transmission}.\hskip 1em plus
  0.5em minus 0.4em\relax Cambridge, UK: Cambridge University Press, 2003.

\bibitem{PK2004}
P.~Koopman and T.~Chakravarty, ``Cyclic redundancy code ({CRC}) polynomial
  selection for embedded networks,'' in \emph{Int. Conf. Dependable Systems and
  Networks}, Jun. 2004, pp. 145--154.

\bibitem{CY2015}
C.~Y. Lou, B.~Daneshrad, and R.~D. Wesel, ``Convolutional-code-specific {CRC}
  code design,'' \emph{{IEEE} Trans. Commun.}, vol.~63, no.~10, pp. 3459--3470,
  Oct. 2015.

\bibitem{Johannesson1999}
R.~Johannesson and K.~S. Zigangirov, \emph{Fundamentals of Convolutional
  Coding}, J.~B. Anderson, Ed.\hskip 1em plus 0.5em minus 0.4em\relax New
  Jersey, USA: IEEE Press, 1999.

\bibitem{SS1994}
N.~Seshadri and C.~E.~W. Sundberg, ``List viterbi decoding algorithms with
  applications,'' \emph{{IEEE} Trans. Commun.}, vol.~42, no. 234, pp. 313--323,
  Feb. 1994.

\bibitem{Soong1991}
F.~K. Soong and E.~F. Huang, ``A tree-trellis based fast search for finding the
  n-best sentence hypotheses in continuous speech recognition,'' in \emph{Proc.
  Int. Conf. Acoustics, Speech, and Signal Processing (ICASSP)}, Apr. 1991, pp.
  705--708 vol.1.

\bibitem{Nill1995}
C.~Nill and C.~E.~W. Sundberg, ``List and soft symbol output viterbi
  algorithms: extensions and comparisons,'' \emph{{IEEE} Trans. Commun.},
  vol.~43, no. 2/3/4, pp. 277--287, Feb. 1995.

\bibitem{Roder2006}
M.~Roder and R.~Hamzaoui, ``Fast tree-trellis list viterbi decoding,''
  \emph{{IEEE} Trans. Commun.}, vol.~54, no.~3, pp. 453--461, Mar. 2006.

\bibitem{Wang2008}
R.~Wang, W.~Zhao, and G.~B. Giannakis, ``{CRC}-assisted error correction in a
  convolutionally coded system,'' \emph{{IEEE} Trans. Commun.}, vol.~56,
  no.~11, pp. 1807--1815, Nov. 2008.

\bibitem{Sybis2016}
M.~Sybis, K.~Wesolowski, K.~Jayasinghe, V.~Venkatasubramanian, and
  V.~Vukadinovic, ``Channel coding for ultra-reliable low-latency communication
  in 5g systems,'' in \emph{2016 IEEE 84th Vehicular Technology Conference
  (VTC-Fall)}, Sept 2016, pp. 1--5.

\bibitem{Chen2001}
B.~Chen and C.~.~W. Sundberg, ``List viterbi algorithms for continuous
  transmission,'' \emph{IEEE Transactions on Communications}, vol.~49, no.~5,
  pp. 784--792, May 2001.

\bibitem{Bai2004}
C.~Bai, B.~Mielczarek, W.~A. Krzymien, and I.~J. Fair, ``Efficient list
  decoding for parallel concatenated convolutional codes,'' in \emph{2004 IEEE
  15th International Symposium on Personal, Indoor and Mobile Radio
  Communications (IEEE Cat. No.04TH8754)}, vol.~4, Sept 2004, pp. 2586--2590
  Vol.4.

\bibitem{Lijofi2004}
L.~Lijofi, D.~Cooper, and B.~Canpolat, ``A reduced complexity list
  single-wrong-turn (swt) viterbi decoding algorithm,'' in \emph{2004 IEEE 15th
  International Symposium on Personal, Indoor and Mobile Radio Communications
  (IEEE Cat. No.04TH8754)}, vol.~1, Sept 2004, pp. 274--279 Vol.1.

\bibitem{Polyanskiy2010}
Y.~Polyanskiy, H.~V. Poor, and S.~Verdu, ``Channel coding rate in the finite
  blocklength regime,'' \emph{IEEE Transactions on Information Theory},
  vol.~56, no.~5, pp. 2307--2359, May 2010.

\bibitem{Segura2018}
J.~Font-Segura, G.~Vazquez-Vilar, A.~Martinez, A.~G. i~Fàbregas, and
  A.~Lancho, ``Saddlepoint approximations of lower and upper bounds to the
  error probability in channel coding,'' in \emph{2018 52nd Annual Conference
  on Information Sciences and Systems (CISS)}, March 2018, pp. 1--6.

\bibitem{ErrorControlCoding}
S.~Lin and D.~J. Costello, \emph{Error Control Coding: fundamentals and
  applications}.\hskip 1em plus 0.5em minus 0.4em\relax New Jersey, USA:
  Pearson Prentice Hall, 2004.

\end{thebibliography}
% that's all folks
\end{document}